\documentclass{amsproc}
\usepackage{amsmath, amsthm, amssymb, graphicx,slashed}
\usepackage{url}
\newtheorem{theorem}{Theorem}[section]

\theoremstyle{definition}

\theoremstyle{remark}
\newtheorem{remark}[theorem]{Remark}

\numberwithin{equation}{section}

\title{Geometric Langlands From Six Dimensions}

\author{Edward Witten}
\address{Theory Group, CERN, Geneva Switzerland.  On leave from
School of Natural Sciences, Institute for Advanced Study, Princeton
NJ 08540 USA. }
\email{witten@ias.edu}
\thanks{Supported in part by NSF Grant Phy-0503584.}

\def\R{{\Bbb{R}}}\def\Z{{\Bbb{Z}}}

\begin{document}
\def\C{\Bbb{C}}
\def\M{{\cal M}}
\def\Bbb{\mathbb}
\def\frak{\mathfrak}
\subjclass{}
\date{May, 2009}

\begin{abstract}
Geometric Langlands duality is usually formulated as a statement
about Riemann surfaces, but it can be naturally understood as a
consequence of electric-magnetic duality of four-dimensional gauge
theory.  This duality in turn is naturally understood as a
consequence of the existence of a certain exotic supersymmetric
conformal field theory in six dimensions.  The same six-dimensional
theory also gives a useful framework for understanding some recent
mathematical results involving a counterpart of geometric Langlands
duality for complex surfaces.  (This article is based on a lecture
at the Raoul Bott celebration, Montreal, June 2008.)
\end{abstract}

\maketitle
\def\D{{\mathcal D}}
\def\Z{{\Bbb Z}}

\section{Introduction}
\def\H{{\mathcal H}}
\def\cal{\mathcal}

A $d$-dimensional quantum field theory (QFT) associates a number,
known as the partition function $Z(X_d)$, to a closed $d$-manifold
$X_d$ endowed with appropriate structure.\footnote{We will slightly
relax the usual axioms in section \ref{confblock}.} Depending on the
type of QFT considered, the requisite structure may be a smooth
structure, a conformal structure, or a Riemannian metric, possibly
together with an orientation or a spin structure, etc.  In physical
language, the partition function can usually be calculated via a
path integral over fields on $X$. However, this lecture will be
partly based on an exception to that statement.

To a closed $d-1$-dimensional manifold $X_{d-1}$ (again with some
suitable structure), a $d$-dimensional QFT associates a vector space
${\cal H}(X_{d-1})$, usually called the space of physical states. In
the case of a unitary QFT (such as the one associated with the
Standard Model of particle physics), ${\cal H}$ is actually a
Hilbert space, not just a vector space. The quantum field theories
considered in this lecture are not necessarily unitary.  The
partition function associated to the empty $d$-manifold is
$Z(\varnothing)=1$, and the vector space associated to the empty
$d-1$-manifold is ${\cal H}(\varnothing)=\C$.

There is a natural link between these structures.   To a
$d$-manifold $X_d$ with boundary $X_{d-1}$, a $d$-dimensional QFT
associates a vector $\psi_{X_d}\in {\cal H}(X_{d-1})$. (In physical
terminology, $\psi_{X_d}$ can usually be computed by performing a
path integral for fields on $X_{d}$ that have prescribed behavior
along its boundary.)  This generalizes the partition function, since
if $X_{d-1}=\varnothing$, then $\psi_{X_d}\in {\cal
H}(\varnothing)=\C$ is simply a complex number, which is the
partition function $Z(X_d)$.

\def\C{{\mathcal C}}
What I have said so far is essentially rather familiar to
physicists.  (The reason for the word ``essentially'' in the last
sentence is that for most physical applications, a less abstract
formulation is adequate.)  Less familiar is that it is possible to
continue the above discussion to lower dimensions.  The next step in
the hierarchy is that to a closed $d-2$-manifold $X_{d-2}$ (with
appropriate structure) one associates a {\it category}
$\C(X_{d-2})$.  Then, for example, to a $d-1$-manifold $X_{d-1}$
with boundary $X_{d-2}$, one associates an object ${\mathcal
P}(X_{d-1})$ in the category $\C(X_{d-2})$.  (For relatively
informal accounts of these matters from different points of view,
see \cite{Freed,BaDo}; for some recent developments, see
\cite{Lurie} as well as \cite{GS}.)

\subsection{Categories And Physics}

In practice, physicists do not usually specify what should be
associated to $X_{d-2}$. This is not necessary for most purposes --
certainly not in standard applications of QFT to particle physics or
condensed matter physics. However, before getting to the main
subject of this talk, I will briefly explain a few cases in which
that language is or might be useful for physicists.

So far, the most striking physical application of the ``third
tier,'' that is the extension of QFT to codimension two, is in
string theory, where one uses two-dimensional QFT to describe the
propagation of a string. In this case, since $d=2$, a $d-2$-manifold
is just a point. So the extra layer of structure is just that the
theory is endowed with a category $\C$, which is the category of
what physicists call boundary conditions in the quantum field
theory, or $D$-branes.

\def\B{{\mathcal B}}
 For $d=2$, a connected $d-1$-manifold with boundary is simply a
closed interval $I$, whose boundary consists of two points. To
define a space ${\mathcal H}(I)$ of physical states of the open
string, one needs boundary conditions $\B$ and $\B'$ at the two ends
of $I$. To emphasize the dependence on the boundary conditions, the
space of physical states is better denoted as ${\mathcal
H}(I;\B,\B')$. In category language, this space of physical states
is called the space of morphisms in the category, ${\rm
Hom}_\C(\B,\B')$.  (This construction has two variants that differ
by whether the manifolds considered are oriented; they are both
relevant to string theory.)

Another case in which the third tier can be usefully invoked, in
practice, is three-dimensional Chern-Simons gauge theory.  This is a
quantum field theory for $d=3$ with a compact gauge group $G$ and a
Lagrangian that is, roughly speaking,\footnote{This formulation
suffices if $G$ is simple, connected, and simply-connected.  In
general, $k$ is an element of $H^4(BG,\Bbb{Z})$.} an integer $k$
times the Chern-Simons functional. A closed $d-2$-manifold is now a
circle, and again, the extra layer of structure is that a category
$\C$ is associated to the theory; it is the category of positive
energy representations of the loop group of $G$ at level $k$.

\def\O{{\mathcal O}}
\def\P{{\mathcal P}}
\def\S{{\mathcal S}}  Finally, the state of the Universe in the presence of a
black hole or a cosmological horizon is sometimes described in terms
of a density matrix rather than an ordinary quantum state, to
account for one's ignorance of what lies beyond the horizon.  This
point of view (which notably has been advocated by Stephen Hawking)
can possibly be usefully reformulated or refined in terms of
categories. The idea here would be that, in $d$-dimensional
spacetime, the horizon of a black hole (or a cosmological horizon)
is a closed $d-2$-manifold. Indeed, suppose that $X_d $ is a
$d$-dimensional Lorentz signature spacetime with an ``initial time''
hypersurface $X_{d-1}$.  Suppose further that a black hole is
present; its horizon intersects $X_{d-1}$ on a codimension two
submanifold $X_{d-2}$.  It is plausible that to $X_{d-2}$, we should
associate a category $\C$, and then to $X_{d-1}$ we would associate
not -- as we would in the absence of the black hole -- a physical
Hilbert space $\H(X_{d-1})$ -- but rather an object $\mathcal P$ in
that category.

To make this more concrete, suppose for example that $\C$ is the
category of representations of an algebra $\S$.  Then $\P$ is an
$\S$-module, which in this context would mean a Hilbert space
$\H(X_{d-1};X_{d-2})$ with an action of $\S$.  Physical operators
would be operators on this Hilbert space that commute with $\S$.
Intuitively, $\S$ is generated by operators that act  behind the
horizon of the black hole.  (That cannot be a precise description in
quantum gravity, where the position of the horizon can fluctuate.)
This point of view is most interesting if the algebra $\S$ is not of
Type I, so that it does not have irreducible modules and the
category of $\S$-modules is not equivalent to the category of vector
spaces.  At any rate, even if the categorical language is relevant
to quantum black holes, it may be oversimplified to suppose that
$\C$ is the category of representations of some algebra.

\subsection{Geometric Langlands}

Our aim here, however, is to understand not black holes but the
geometric Langlands correspondence.  In this subject, one studies a
Riemann surface $C$, but the basic statements that one makes are
about {\it categories} associated to $C$.  Indeed, the basic
statement is that two categories associated to $C$ are equivalent to
each other.

For $G$ a simple complex Lie group, let $Y_G(C)={\mathrm
{Hom}}(\pi_1(C),G)$ be the moduli stack of flat $G$-bundles over
$C$.  And let $Z_G(C)$ be the moduli stack of holomorphic
$G$-bundles over $C$.

\def\cD{\mathcal D}
To the group $G$, we associate its Langlands \cite{L} or GNO
\cite{GNO} dual group $G^\vee$.  (The root lattice of $G$ is the
coroot lattice of $G^\vee$, and vice-versa.)  Then the basic
assertion of the geometric Langlands correspondence \cite{BD} is
that the category of coherent sheaves on $Y_{G^\vee}(C)$ is
naturally equivalent to the category of ${\cD}$-modules on $Z_G(C)$.

If we are going to interpret this statement in the context of
quantum field theory, we should start with a theory in dimension
$d=4$, so that it will associate a category to a manifold of
dimension $d-2=2$, in this case the two-manifold $C$. We need then
an equivalence between a quantum field theory defined using $G$ and
a quantum field theory defined using $G^\vee$, both in four
dimensions.  In fact, there is a completely canonical theory with
the right properties.  It is the maximally supersymmetric Yang-Mills
theory in four dimensions.

This theory, which has ${\mathcal N}=4$ supersymmetry, depends on
the choice of a compact\footnote{In the formulation via gauge
theory, we begin with a compact gauge group, whose complexification
then naturally appears by the time one makes contact with the usual
statements about geometric Langlands.  Geometric Langlands is
usually described in terms of this complexification.} gauge group
$G$. It also depends on the choice of a complex-valued  quadratic
form on the Lie algebra $\frak g$ of $G$; the imaginary part of this
quadratic form is required to be positive definite. If $G$ is
simple, then Lie theory lets us define a natural invariant quadratic
form on $\frak g$ (short coroots have length squared 2), and any
such form is a complex multiple of this one.  We write the multiple
as
\begin{equation}\label{golfo}\tau=\frac{\theta}{2\pi}+\frac{4\pi
i}{e^2},\end{equation} where $e$ and $\theta$ (known as the gauge
coupling constant and theta-angle) are real. We call $\tau$ the
coupling parameter.

The classic statement  (which evolved from early ideas of Montonen
and Olive \cite{MO}) is that ${\mathcal N}=4$ super Yang-Mills
theory with gauge group $G$ and coupling parameter $\tau$ is
equivalent to the same theory with dual gauge group $G^\vee $ and
coupling parameter \begin{equation}\label{gulp}\tau^\vee=-1/n_{\frak
g}\tau.\end{equation}
 (Here $n_{\frak g}$ is the ratio of length
squared of long and short roots of $G$ or $G^\vee$.)  The
equivalence between the two theories exchanges electric and magnetic
fields, in a suitable sense, and is known as electric-magnetic
duality.  There are also equivalences under
\begin{equation}\label{hombo}\tau\to\tau+1,~\tau^\vee\to\tau^\vee+1,\end{equation}  that can be
seen semiclassically (as a reflection of the fact that the instanton
number of a classical gauge field is integer-valued). The
non-classical equivalence (\ref{gulp}) combines with the
semiclassical equivalences (\ref{hombo}) to an infinite discrete
structure.  For instance, if $G$ is simply-laced, then $n_{\frak
g}=1$, $G$ and $G^\vee$ have the same Lie algebra, for many purposes
one can ignore the distinction between $\tau$ and $\tau^\vee$, and
the symmetries (\ref{gulp}) and (\ref{hombo}) generate an action of
the infinite discrete group $SL(2,\Z)$ on $\tau$.

There is a ``twisting procedure'' to construct topological quantum
field theories (TQFT's) from physical ones.  Applied to ${\mathcal
N}=2$ super Yang-Mills theory, this procedure leads to Donaldson
theory of smooth four-manifolds.  Applied to ${\mathcal N}=4$ super
Yang-Mills theory, the twisting procedure leads to three possible
constructions.  Two of these are quite similar to Donaldson theory
in their content, while the third is related to geometric Langlands
\cite{KW}.

The equivalence between this third twisting for the two groups $G$
and $G^\vee$ (and with an inversion of the coupling parameter) leads
precisely at the level of ``categories,'' that is for two-manifolds,
to the geometric Langlands correspondence.  (The underlying
electric-magnetic duality treats $G$ and $G^\vee$ symmetrically. But
the twisting depends on a complex parameter; the choice of this
parameter breaks the symmetry between $G$ and $G^\vee$.  That is why
the usual statement of the geometric Langlands correspondence treats
$G$ and $G^\vee$ asymmetrically.)

So this is the basic reason that geometric Langlands duality, most
commonly understood as a statement about Riemann surfaces, arises
from a quantum field theory in {\it four} dimensions.

\remark For another explanation of why four dimensions is a natural
starting point for geometric Langlands, see \cite{W1}.  This
explanation uses the fact that the mathematical theory as usually
developed is based on moduli stacks rather than moduli spaces; but a
two-dimensional sigma model whose target is the moduli stack of
bundles is best understood as a four-dimensional gauge theory. This
relies on the gauge theory interpretation of the moduli stack,
introduced in a well-known paper by Atiyah and Bott \cite{AB}.

\section{Defects Of Various Dimension}\label{defect}

In the title of this talk, I promised to get up to six dimensions,
not just four.  Eventually we will, but first we will survey the
role of structures of different dimension in a four-manifold.

\def\O{{\mathcal O}}
Suppose that a quantum field theory on a manifold $M$ is defined by
some sort of path integral, schematically
\begin{equation}\label{golf}\int DA\dots \exp\left(-\int_M
L\right),\end{equation} where $L$ is a Lagrangian density that
depends on some fields $A$ (and perhaps on additional fields that
are not written). ``Inserting a local operator $\O(p)$ at a point
$p\in M$'' means modifying the path integral at that point. This may
be done by including a factor in the path integral that depends on
the fields and their derivatives only at $p$.  It may also be done
in some more exotic way, such as by prescribing a singularity that
the fields should have near $p$.

In addition to local operators, we can also consider modifications
of the theory that are supported on a $p$-dimensional submanifold
$N\subset M$. We give some examples shortly.  A local operator is
the case $p=0$.  The general case we call a $p$-manifold operator.

In much of physics, the important operators are local operators.
This is also the case in Donaldson theory.  The local operators that
are important in Donaldson theory are related to characteristic
classes of the universal bundle.

I should point out that geometrically, a local operator may be a
tensor field of some sort on $M$; it may be, for example, a $q$-form
for some $q$.  If $\O_q$ is a local operator valued in $q$-forms, we
can integrate it over a $q$-cycle $W_q\subset M$ to get
$\int_{W_q}\O_q$.  The most important operators in Donaldson theory
are of this kind, with $q=2$.  For our purposes, we need not
distinguish a local operator from such an integral of one.  (What we
call a $p$-manifold operator cannot be expressed as an integral of
$q$-manifold operators with $q<p$.)

Local operators also play a role in geometric Langlands.  Indeed, a
construction analogous to that of Donaldson is relevant.  Imitating
the construction of Donaldson theory and then applying
electric-magnetic duality, one arrives at results, many of which are
known in the mathematical literature, comparing group theory of $G$
to cohomology of certain orbits in the affine Grassmannian of
$G^\vee$.

But local operators are not the whole story.  In gauge theory, for
example, given an oriented circle $S\subset M$, and a representation
$R$ of $G$, we can form the trace of the holonomy of the connection
$A$ around $S$ in the given representation.  Physicists denote this
as
\begin{equation}\label{cithol}W_R(S)={\rm
Tr}_R\,P\exp\left(-\oint_SA\right).\end{equation} When included as a
factor in a quantum path integral, $W_R(S)$ is known as a Wilson
operator. Wilson operators were introduced over thirty years ago in
formulating a criterion for quark confinement in the theory of the
strong interactions.

$W_R(S)$ cannot be expressed as the integral over $S$ of a local
operator.  We call it a one-manifold operator.

Electric-magnetic duality inevitably converts $W_R(S)$ to another
one-manifold operator, which was described by 't Hooft in the late
1970's.  The 't Hooft operator is defined by prescribing a
singularity that the fields should have along $S$.  (See \cite{KW}
for a review.  Operators defined in this way are often called
disorder operators, while operators like the Wilson operator that
are defined by interpreting a classical expression in quantum
mechanics are called order operators.)  The possible singularities
in $G$ gauge theory are in natural correspondence with
representations $R^\vee$ of the dual group $G^\vee$.
Electric-magnetic duality maps a Wilson operator in $G^\vee$ gauge
theory associated with a representation $R^\vee$ to an 't Hooft
operator in $G$ gauge theory that is also associated with $R^\vee$.

If one specializes to the situation usually studied in the geometric
Langlands correspondence, the 't Hooft operators correspond to the
usual geometric Hecke operators of that subject.  The
electric-magnetic duality between Wilson and 't Hooft operators
leads to the usual statement that a coherent sheaf on
$Y_{G^\vee}(C)$ that is supported at a point is dual to a Hecke
eigensheaf on $Z_G(C)$.  (Saying that a $\D$-module on $Z_G(C)$ is a
Hecke eigensheaf is the geometric analog of saying that a classical
modular form is a Hecke eigenform.)

Moving up the chain, the next step is a two-manifold operator.  In
general, in $d$-dimensional gauge theory, one can define a
$d-2$-manifold operator as follows.  One omits from $M$ a
codimension two submanifold $L$.  Then, fixing a conjugacy class in
$G$, one considers gauge fields on $M\backslash L$ with holonomy
around $L$ in the prescribed conjugacy class.

For $d=4$, we have $d-2=2$, so $L$ is a two-manifold.  Classical
gauge theory in the presence of a singularity of this kind has been
studied in the context of Donaldson theory by Kronheimer and Mrowka.
In geometric Langlands, to get a class of two-manifold operators
that is invariant under electric-magnetic duality, one must
incorporate certain quantum parameters in addition to the holonomy
\cite{GW}. Once one does this, one gets a natural quantum field
theory framework for understanding ``ramification,'' i.e. the
geometric Langlands analog of ramification in number theory.

The next case are operators supported on a three-manifold $W\subset
M$. With $M$ being of dimension four, $W$ is of codimension one and
locally divides $M$ into two pieces.  The theory of such
three-manifold operators is extremely rich and \cite{GaW1,GaW} there
are many interesting constructions, even if one requires that they
should preserve the maximum possible amount of supersymmetry (half
of the supersymmetry).

For example, the gauge group can jump in crossing $W$.  We may have
$G$ gauge theory one side and $H$ gauge theory on the other.  If $H$
is a subgroup of $G$, a construction is possible that is related to
what Langlands calls functoriality.  Other universal constructions
of geometric Langlands -- including the universal kernel that
implements the duality -- are similarly related to supersymmetric
three-manifold operators.

As long as we are in four dimensions, this is the end of the road
for modifying a theory on a submanifold.  A modification in four
dimensions would just mean studying a different theory. So to
continue the lecture, we will, as promised in the title, try to
relate geometric Langlands to a phenomenon above four dimensions.

\section{Selfdual Gerbe Theory In Six Dimensions}\label{gerbeth}

Until relatively recently, it was believed that four was the maximum
dimension for nontrivial (nonlinear or non-Gaussian) quantum field
theory. One of the surprising developments coming from string theory
is that nontrivial quantum field theories exist up to (at least) six
dimensions.

\def\d{{\mathrm d}}
To set the stage, I will begin by sketching a linear, but subtle,
quantum field theory in six dimensions.  The nonlinear case is
discussed in section \ref{nonca}.

In six dimensions, with Lorentz signature $-+++++$, a real
three-form $H$ can be selfdual, obeying $H=\star H$, where $\star$
is the Hodge star operator.\footnote{\label{euclid} The quantum
theory of a real selfdual threeform in six dimensions can be
analytically continued to Euclidean signature, whereupon $H$ is
still selfdual but is no longer real. Such a continuation will be
made later. In general, analytic continuation from Lorentz to
Euclidean signature and back is an important tool in quantum field
theory; the basic reason that it is possible is that in Lorentz
signature  the energy is non-negative.} Let us consider such an $H$
and endow it with a hyperbolic equation of motion
\begin{equation}\label{endow}{\d} H = 0.\end{equation}
That equation is analogous to the Bianchi identity $\d F=0$ for the
curvature two-form $F$ of a line bundle.  It means that (in a
mathematical language that physicists generally do not use) $H$ can
be interpreted as the curvature of a $U(1)$ gerbe with connection.

In contrast to gauge theory, there is no way to derive this system
from an action.  The natural candidate for an action, on a
six-manifold $M_6$, would seem to be $\int_{M_6}H\wedge\star H$, but
if $H$ is self-dual this is the same as $\int_{M_6}H\wedge H=0$.

Nevertheless, there is a quantum field theory of the closed,
selfdual $H$ field.  To explain how one part of the structure of
quantum field theory emerges, suppose that the Lorentz signature
six-manifold $M_6$ admits a global Cauchy hypersurface $M_5$.  $M_5$
is thus a five-dimensional Riemannian manifold.  Fixing the
topological type of a $U(1)$ gerbe in a neighborhood of $M_5$, the
space of gerbe connections with selfdual curvature, modulo gauge
transformations, is an (infinite-dimensional) symplectic manifold in
a natural way. (Roughly speaking, if $B$ is the gerbe connection,
then the symplectic form is defined by the formula
$\omega=\int_{M_5}\delta B\wedge\d\delta B$.)  Quantizing this
space, we get a Hilbert space associated to $M_5$.  This association
of a Hilbert space to a five-manifold is part of the usual data of a
six-dimensional quantum field theory. The rest of the structure can
also be found, with some effort. (For a little more detail, see
\cite{EW,HenningsonA,HenningsonB}.)

An important fact is that the quantum field theory of the $H$ field
is conformally invariant.  Classically, the equations $H=\star H$,
$\d H=0$, are conformally invariant.  The passage to quantum
mechanics preserves this property, because the theory is linear.

\def\Z{{\Bbb Z}}
Now let us consider the special case that our
six-manifold\footnote{Henceforth, and until section \ref{compac}, we
generally work in Euclidean signature, using the analytic
continuation mentioned in footnote \ref{euclid}.} takes the form
$M_6=M_4\times T^2$, where $M_4$ is a four-manifold and $T^2$ is a
two-torus.  We assume a product conformal structure on $M_4\times
T^2$.  After making a conformal rescaling to put the metric on $T^2$
in a standard form (say a flat metric of unit area), we are left
with a Riemannian metric on $M^4$. The conformal structure of $T^2$
is determined by the choice of a point $\tau$ in the upper half of
the complex plane -- modulo the action of $SL(2,\Z)$.

Next in $M_4\times T^2$, let us keep fixed the second factor, with a
definite metric, and let the first factor vary.  We let $M_4$ be an
arbitrary four-manifold with boundaries, corners, etc. Starting with
a conformal field theory on $M_6$, this process gives us a
four-dimensional quantum field theory (not conformally invariant)
that depends on $\tau$ as a parameter.  Clearly, the induced
four-dimensional theory depends on the conformal structure of $T^2$
only up to isomorphism.  So if we parametrize the induced
four-dimensional theory by $\tau$, we will have a symmetry under the
action of $SL(2,\Z)$ on $\tau$.

\def\CC{\Bbb{C}}
The induced four-dimensional quantum field theory is actually
closely related to $U(1)$ gauge theory, which is its ``infrared
limit.''  Let us think of $T^2$ as $\CC/\Lambda$, where $\CC$ is the
complex plane parametrized by $z=x+iy$ and $\Lambda$ is the lattice
generated by complex numbers 1 and $\tau$.  Further, make an ansatz
\begin{equation}\label{hhh}H=F\wedge \d x+\star F \wedge \d y,\end{equation}
 where $F$ is a two-form on
$M_4$ (pulled back to $M_6=M_4\times T^2$), and $\star$ is the
four-dimensional Hodge star operator.  Then the equations $\d H=0$
become Maxwell's equations
\begin{equation}\label{stromb} \d F=\d \star F = 0.\end{equation}

This gives an embedding of four-dimensional $U(1)$ gauge theory in
the six-dimensional theory.  To be more precise, we should think of
$H$ as the curvature of a $U(1)$ gerbe connection; then $F$ is the
curvature of a $U(1)$ connection.  Of course, we have described the
embedding classically, but it also works quantum mechanically.

This construction is more than an embedding of four-dimensional
$U(1)$ gauge theory in a six-dimensional theory.  The
four-dimensional $U(1)$ gauge theory is the infrared limit of the
six-dimensional theory in the following sense.  We have endowed
$M_6$ with a product metric $g_6$ that we can write schematically as
$g_6=g_4\oplus g_2$, where $g_4$ and $g_2$ are metrics on $M_4$ and
$T^2$, respectively.  Now we modify $g_6$ to $g_6(t)=t^2g_4\oplus
g_2$, where $t$ is a real parameter.  The claim is that for
$t\to\infty$, the theory on $M_6$ converges to $U(1)$ gauge theory
on $M_4$.  (This theory is conformally invariant, so the $t^2$
factor in the metric of $M_4$ can be dropped.) This is usually
described more briefly by saying that $U(1)$ gauge theory on $M_4$
is the long distance or infrared limit of the underlying theory on
$M_6$.

Even though $U(1)$ gauge theory on $M_4$ gives an effective and
useful description of the large $t$ limit of the six-dimensional
theory on $M_6$, something is obscured in this description.  The
process of compactifying on $T^2$ and taking the large $t$ limit is
canonical in that it depends only on the geometry of $T^2$ and not
on a choice of coordinates.  But to go to a description by $U(1)$
gauge theory, we used the ansatz (\ref{hhh}), which depended on a
choice of coordinates $x$ and $y$.  As a result, some of the
underlying symmetry is hidden in the description by $U(1)$ gauge
theory.

Concretely, though the  six-dimensional theory does not have a
Lagrangian, the four-dimensional $U(1)$ gauge theory does have one:
\begin{equation}\label{lagr}I=\frac{1}{4e^2}\int_{M_4}F\wedge \star
F+\frac{\theta}{8\pi^2}\int F\wedge F.\end{equation} The coupling
parameter
\begin{equation}\label{gomo}\tau=\frac{\theta}{2\pi}+\frac{4\pi
i}{e^2}.\end{equation} of the abelian gauge theory is simply the
$\tau$-parameter of the $T^2$ in the underlying six-dimensional
description.

The six-dimensional theory depends on $\tau$ only modulo the usual
 $SL(2,\Z)$ equivalence $\tau\to (a\tau+b)/(c\tau+d)$, with
integers $a,b,c,d$ obeying $ad-bc=1$, since values of $\tau$ that
differ by the action of $SL(2,\Z)$ correspond to equivalent tori.
Therefore, the limiting four-dimensional $U(1)$ gauge theory must
also have $SL(2,\Z)$ symmetry.  However, there is no such classical
symmetry. Manifest $SL(2,\Z)$ symmetry was lost in the reduction
from six to four dimensions, because the ansatz (\ref{hhh}), which
was the key step in reducing to four dimensions, is not
$SL(2,\Z)$-invariant. Hence this ansatz leads to a four-dimensional
theory with a ``hidden'' $SL(2,\Z)$ symmetry, one which relates the
description by a $U(1)$ gauge field with curvature $F$ to a
different description by a different $U(1)$ gauge field with another
curvature form (which, roughly speaking, is related to $F$ by the
action of $SL(2,\Z)$).

What we get this way is an $SL(2,\Z)$ symmetry of quantum $U(1)$
gauge theory that does not arise from a symmetry of the classical
theory.  To physicists, this symmetry is known as electric-magnetic
duality.  The name is motivated by the fact that an exchange
$(x,y)\to (y,-x)$ in (\ref{hhh}), which is a special case of
$SL(2,\Z)$, would exchange $F$ and $\star F$, and thus in
nonrelativistic terminology would exchange electric and magnetic
fields.

So we have seen that   electric-magnetic duality in $U(1)$ gauge
theory in four dimensions follows from the existence of a suitable
conformal field theory in six dimensions \cite{V}.  The starting
point in this particularly nice explanation is the existence in six
dimensions of a quantum theory of a gerbe with selfdual curvature.
(It is also possible to demonstrate the four-dimensional duality by
a direct calculation, involving a sort of Fourier transform in field
space; see \cite{Witten:1995gf}.)

\section{The Nonabelian Case}\label{nonca}

Since there is not a good notion classically of a gerbe whose
structure group is a simple nonabelian Lie group, one might think
that it is too optimistic to look for an analogous explanation of
electric-magnetic duality for nonabelian groups.  However, it turns
out that such an explanation does exist -- in the maximally
supersymmetric case.

The picture is simplest to describe if $G$ is simply-laced, in which
case $G$ and $G^\vee$ have the same Lie algebra (and to begin with,
we will ignore the difference between them, though this is precisely
correct only if $G=E_8$; a more complete picture can be found in
section \ref{confblock}). For $G$ to be simply-laced is equivalent
to the condition that $n_{\frak g}=1$ in eqn. (\ref{gulp}). For many
purposes, we can ignore the difference between $\tau$ and
$\tau^\vee$, and then the quantum duality (\ref{gulp}) and the
semiclassical equivalence (\ref{hombo}) combine to an action of
$SL(2,\Z)$ on $\tau$.

For every simply-laced Lie group $G$, there is a six-dimensional
conformal field theory that in some sense is associated with gerbes
of type $G$.  The theory is highly supersymmetric, so supersymmetry
is essential in what follows.  The existence of this theory was
discovered in string theory in the mid-1990's. (The first hint
\cite{Wittenz} came by considering Type IIB superstring theory at an
ADE singularity.) Its existence is probably our best explanation of
electric-magnetic duality -- and therefore, in particular, of
geometric Langlands duality.  It is, in the jargon of quantum field
theory, an isolated, non-Gaussian conformal field theory. This means
among other things that it cannot be properly described in terms of
classical notions such as partial differential equations.

However, it has two basic properties which in a sense
 justify thinking of it as a quantum theory of nonabelian gerbes.
Each property involves a perturbation of some kind that causes a
simplification to
   a theory that {\it can} be given a classical description.
The two perturbations are as follows:

(1) After a perturbation in the vacuum expectation values of certain
fields (which are analogous to the conjectured Higgs field of
particle physics), the theory reduces at low energies to a theory of
gerbes, with selfdual curvature, and structure group the maximal
torus $T$ of $G$. This notion does make sense classically, since $T$
is abelian. In fact, the selfdual gerbe theory of $T$ is much like
the $U(1)$ theory described in Section 3, with $U(1)$ replaced by
$T$. (Supersymmetry plays a fairly minor role in the abelian case.)

\def\Tr{{\rm Tr}}
(2) Let $M_6=M_5\times S^1$ be the product of a five-manifold $M_5$
with a circle; we endow it with a product metric $g_6=g_5\oplus
g_1$. The six-dimensional theory on $M_6$ has a description (valid
at long wavelengths) in terms of $G$ gauge fields (and other fields
related to them by supersymmetry) on $M_5$, but this description
involves a highly nonclassical trick. If the circle factor of
$M_6=M_5\times S^1$ has radius $R$, then the effective action for
the gauge fields in five dimensions is {\it inversely} proportional
to $R$:
\begin{equation}\label{elfs} I_5=\frac{1}{8\pi R}\int_{M_5}\Tr\,F\wedge
\star F.  \end{equation} The factor of $R^{-1}$ multiplying the
action is a simple consequence of conformal invariance in six
dimensions.  (Under multiplication of the metric of $M_6$ by a
positive constant $t^2$, the Hodge operator $\star$ mapping
two-forms to three-forms in five dimensions is multiplied by $t$,
while $R$ is also multiplied by $t$, so the action in (\ref{elfs})
is invariant.)  Though easily understood, this result is highly
nonclassical.  Eqn. (\ref{elfs}) is a classical Lagrangian for gauge
fields in five dimensions.  Can it arise from a classical Lagrangian
for gauge fields  on $M_6=M_5\times S^1$?  Given a six-dimensional
Lagrangian for gauge fields, we would reduce to a five-dimensional
Lagrangian (for fields that are pulled back from $M_5$) by
integrating over the fibers of the projection $M_5\times S^1\to
M_5$. This would give a factor of $R$ multiplying the
five-dimensional action, not $R^{-1}$. So a theory that leads to the
effective action (\ref{elfs}) cannot arise in this way. The theory
in six dimensions should be, in some sense, not a gauge theory but a
gerbe theory instead, but this does not exist classically in the
nonabelian case.

What I have said so far is that the same six-dimensional quantum
field theory can be simplified to either (i) a six-dimensional
theory of abelian gerbes, or (ii) a five-dimensional theory with a
simple non-abelian gauge group.  The two statements together show
that one cannot do justice to this theory in terms of either gauge
fields (as opposed to gerbes) or abelian groups (as opposed to
non-abelian ones).

\def\tilde{\widetilde}
Now let us look more closely at the implications of the peculiar
factor of $1/R$ in (\ref{elfs}).  We will study what happens for
$M_6=M_4\times T^2$, the same decomposition that we used in studying
the abelian gerbe theory in section \ref{gerbeth}.  However, for
simplicity we will take $T^2= S^1\times \tilde S^1$ to be the
orthogonal product of a circle $S^1$ of radius $R$ and a second
circle $\tilde S^1$ of radius $S$. The tau parameter of such a torus
(which is made by identifying the sides of a rectangle of height and
width $2\pi R$ and $2\pi S$) is
\begin{equation}\label{zork}\tau=i\frac{S}{R}~~{\rm
or}~~\tau=i\frac{R}{S},\end{equation} depending on how one
identifies the rectangle with a standard one. The two values of
$\tau$ differ by
\begin{equation}\label{zoro}\tau\to -\frac{1}{\tau}.\end{equation}

We first view the six-manifold $M_6$ as $M_6=M_5\times S^1$, where
$M_5=M_4\times \tilde S^1$. The six-dimensional theory on $M_6$
reduces at long distances to a supersymmetric gauge theory on $M_5$.
According to (\ref{elfs}), the action for the gauge fields is
\begin{equation}\label{zelfs} I_5=\frac{1}{8\pi R}\int_{M_4\times \tilde S^1}\Tr\,F\wedge
\star F.  \end{equation} Now if $M_4$ is much larger than $\tilde
S^1$, then at long distances we can assume that the fields are
invariant under rotation of $\tilde S^1$ and we can deduce an
effective action in four dimensions by integration over the fiber of
the projection $M_4\times \tilde S^1\to M_4$.   This second step is
purely classical, so it gives a factor of $S$.  The effective action
in four dimensions is thus
\begin{equation}\label{elfsof} I_4=\frac{S}{8\pi R}\int_{M_4}\Tr\,F\wedge
\star F.  \end{equation}

The important point is that this formula is not symmetric in $S$ and
$R$, even though they enter symmetrically in the starting point
$M_6=M_4\times S^1\times \tilde S^1$.  Had we exchanged the two
circles before beginning this procedure, we would have arrived at
the same formula for the four-dimensional effective action, but with
$S/R$ replaced by $R/S$.

Looking back to (\ref{zork}) and (\ref{zoro}), we see that the two
formulas differ by $\tau\to -1/\tau$.  Thus, we have
deduced\footnote{To keep the derivation simple, we considered only a
rectangular torus.  It is possible by using eqn. (\ref{omelf}) to
similarly analyze the case of a general torus.} that for
simply-laced $G$, the four-dimensional gauge theory that corresponds
to the maximally supersymmetric completion of (\ref{elfsof}) has a
quantum symmetry that acts on the coupling parameter by $\tau\to
-1/\tau$. This is the electric-magnetic duality that has many
applications in physics and also underlies geometric Langlands
duality.  What we have gained is a better understanding of why it is
true in the nonabelian case.

\remark\label{delf} Unfortunately, despite its importance, there is
no illuminating and widely used name for the six-dimensional QFT
whose existence underlies duality in this way. According to Nahm's
theorem \cite{N}, the superconformal symmetry group of a
superconformal field theory in six dimensions, when formulated in
Minkowski spacetime, is $OSp(2,6|2r)$ for some $r$. The known
examples have $r=1$ or $2$, and the theory with the properties that
I have just described is the ``maximally symmetric'' one with $r=2$.
This theory is rather inelegantly called the six-dimensional $(0,2)$
model of type $G$, where 2 is the value of $r$ (and the
redundant-looking number 0 involves a comparison to six-dimensional
models that are supersymmetric but not conformal).

\remark\label{zorkox}  The bosonic subgroup of $OSp(2,6|2r)$ is
$SO(2,6)\times Sp(2r)$, where $SO(2,6)$ is the conformal group in
six dimensions, and $Sp(2r)$ is an ``internal symmetry group'' (it
acts trivially on spacetime) and is known as the $R$-symmetry group.
Thus, the $R$-symmetry group of the $(0,2)$ model is $Sp(4)$. This
is the group (sometimes called $Sp(2)$) of $2\times 2$ unitary
matrices of quaternions; its fundamental representation is of
quaternionic dimension 2, complex dimension 4, or real dimension 8.
$Sp(4)$ is also the group that acts on the cohomology of a
hyper-Kahler manifold; this is no coincidence, as we will see later.

\subsection{The Space Of Conformal Blocks}\label{confblock}

\def\ZZ{{\mathcal Z}}
Among the simple Lie groups, only $E_8$ is simply-connected and has
a trivial center.  Equivalently, its root lattice $\Gamma$ endowed
with the usual quadratic form is unimodular, that is, equal to its
dual $\Gamma^\vee$.   In general, if $G$ is simple, simply-laced,
and simply-connected, its center is $\ZZ=\Gamma^\vee/\Gamma$, and
the quadratic form on $\Gamma$ leads to a perfect pairing
\begin{equation}\label{perf}\ZZ\times \ZZ\to
\R/\Z=U(1).\end{equation} (The pairing actually takes values in the
subgroup $\Z_n$ of $U(1)$, where $n$ is the smallest integer that
annihilates $\ZZ$.)

\def\R{{\mathcal R}}
What has been said so far is sufficient for $E_8$, but more
generally, a refinement is necessary. (Most of this article does not
depend on the following details.) For orientation, consider
two-dimensional current algebra (that is, the holomorphic part of
the WZW model\footnote{\label{able} There is no satisfactory
terminology in general use. The WZW model is really \cite{WZW} a
two-dimensional quantum field theory that is modular-invariant but
neither holomorphic nor antiholomorphic.  Its holomorphic part
corresponds to what physicists know as two-dimensional current
algebra (which is a much older construction than the WZW model). But
the phrase ``two-dimensional current algebra'' is not well-known to
mathematicians, and may even be unclear nowadays to physicists.}) of
the simply-connected and simply-laced group $G$ at level 1.   This
theory, formulated on a closed Riemann surface $W$, does not have a
unique partition function (which is required in the usual axioms of
quantum field theory, as indicated in the introduction to this
article). Rather, it has a vector space of possible partition
functions, known as the space of conformal blocks.  This vector
space (in the particular case of a simply-laced group $G$ at level
1) can be constructed as follows. The pairing (\ref{perf}) together
with the intersection pairing on the cohomology of $W$ leads to a
perfect pairing
\begin{equation}\label{myp} H^1(W,\ZZ)\times H^1(W,\ZZ)\to U(1).\end{equation}  This pairing enables us
to define a Heisenberg group extension
\begin{equation}\label{tolmex}1\to U(1)\to F\to H^1(W,\ZZ)\to
0.\end{equation} Up to isomorphism, the group $F$ has a single
faithful irreducible representation $\R$ in which $U(1)$ acts in the
natural way; it is obtained by ``quantizing'' the finite group
$H^1(W,\ZZ)$. One picks a decomposition of $H^1(W,\ZZ)$ as $A\times
B$, where $A$ and $B$ (which can be constructed using a system of
$A$-cycles and $B$-cycles on $W$) are maximal subgroups on which the
extension (\ref{tolmex}) is trivial. One then lets $B$ act by
multiplication -- in the sense that $\R$ is the direct sum of all
one-dimensional characters of $B$. Since (\ref{myp}) restricts to a
perfect pairing $A\times B\to U(1)$,
 characters
of $B$ correspond to elements of $A$.  Thus, $\R$ has a unitary
basis consisting of elements $\psi_a$, $a\in A$; the action of $A$
is $a(\psi_{a'})=\psi_{aa'}$, while $B$ acts by $b\psi_a=\exp(2\pi
i(b,a))\psi_a$ (where $\exp(2\pi i(b,a))$ denotes the pairing
between $A$ and $B$).
 The dimension of $\R$ is thus $(\#
\ZZ)^g$, where $g$ is the genus of $W$ and $\# \ZZ$ is the order of
$\ZZ$.

\def\T{{\mathcal T}}
The space of conformal blocks of the level 1 holomorphic WZW model
on a Riemann surface $W$ with a simple and simply-laced symmetry
group $G$ is isomorphic to $\R$. Thus, for $G\not= E_8$, the space
of conformal blocks has dimension bigger than 1.  That means that
this theory does not have a distinguished partition function and so
does not quite obey the full axioms of quantum field theory. One may
either relax the axioms slightly, study the ordinary
(non-holomorphic) WZW model, or in some other way include
holomorphic or non-holomorphic degrees of freedom so as to be able
to define a distinguished partition function.

 The situation in the
six-dimensional $(0,2)$ theory  is similar, with the finite group
$H^3(M_6,\ZZ)$ playing the role of $H^1(W,\ZZ)$ in two dimensions.
{}From (\ref{perf}) and Poincar\'e duality, we again have a perfect
pairing $H^3(M_6,\ZZ)\times H^3(M_6,\ZZ)\to U(1)$, leading to a
Heisenberg group extension
\begin{equation}\label{tolme}1\to U(1)\to F\to H^3(M_6,\ZZ)\to
0.\end{equation} Again, up to isomorphism, $F$ has a unique faithful
irreducible module $\T$ with natural action of $U(1)$. The theory on
a general six-manifold has a space of conformal blocks that is
isomorphic to $\T$. For $G$ a simple and simply-laced Lie group that
is not of type $E_8$, this again represents a slight departure from
the usual axioms of quantum field theory. Our options are analogous
to what they were in the two-dimensional case: live with it (which
will be our choice in the present paper) or consider various more
elaborate constructions in which one can avoid the problem.

Now let us consider an illuminating example.  We take $M_6=M_5\times
S^1$.  We have a decomposition $H^3(M_6,\ZZ)=H^2(M_5,\ZZ)\oplus
H^3(M_5,\ZZ)$.  Calling the summands $A$ and $B$, we can as above
construct the space $\T$ of conformal blocks as the direct sum of
characters of $B$.  Hence, as in the two-dimensional case, $\T$ has
a basis consisting of elements $\psi_a$, $a\in A=H^2(M_5,\ZZ)$.

\def\Gad{G_{\mathrm {a\negthinspace d}}}
On the other hand, the $(0,2)$ model on $M_6=M_5\times S^1$ is
supposed to be related to gauge theory on $M_5$.  So in gauge theory
on $M_5$, we should find a way to define a partition function for
every $a\in H^2(M_5,\ZZ)$.  This is easily done once one appreciates
that one should use the adjoint form of the group, which we will
call $\Gad$. A $\Gad$ bundle over any space $X$ has a characteristic
class $a\in H^2(X,\ZZ)$ (where $\ZZ$ is the center of the
simply-connected group $G$ or equivalently the fundamental group of
$\Gad$).  In $\Gad$ gauge theory on $M_5$, we define for every $a\in
H^2(M_5,\ZZ)$ a corresponding partition function $Z_a$ by summing
the path integral of the theory over all bundles whose
characteristic class equals $a$.

In defining the $Z_a$, we are relaxing the usual  axioms of quantum
field theory a little bit.  If the gauge group is supposed to be
$G$, the characteristic class must vanish and the partition function
is essentially $Z_0$.  (I will omit some elementary factors
involving the order of $\ZZ$.)  If the gauge group is supposed to be
$\Gad$, all values of the characteristic class are allowed and the
partition function is $\sum_a Z_a$.  For groups intermediate between
$G$ and $\Gad$, certain formulas intermediate between those two will
arise.  But for no choice of gauge group is the partition function
precisely $Z_a$, for some fixed and nonzero $a$.  Clearly, on the
other hand, it is natural to permit ourselves to study these
functions.  So this is a situation in which we probably want to be
willing to slightly generalize the usual axioms of quantum field
theory.

Now as before let us consider the case $M_5=M_4\times\tilde S^1$,
where $\tilde S^1$ is another circle, so that $M_6=M_4\times
S^1\times \tilde S^1$ can be viewed in more than one way as the
product of a circle and a five-manifold. For simplicity, let us
assume that $H^1(M_4,\ZZ)=H^3(M_4,\ZZ)=0$. Then
$H^3(M_6,\ZZ)=A\oplus B$, where
\begin{equation}\label{yerto} A=H^2(M_4,\ZZ)\otimes
H^1(S^1,\Z), ~~B= H^2(M_4,\ZZ)\otimes H^1(\tilde
S^1,\Z).\end{equation} The extension is trivial on both $A$ and $B$.
Reasoning as above, the space $\T$ of conformal blocks has a basis
$\psi_a,~a\in A$.  On the other hand, exchanging the roles of $A$
and $B$, it has a second basis $\tilde\psi_b,~b\in B$. As is usual
in quantization, the relation between these two bases (which are
analogous to ``position space'' and ``momentum space'') is given by
a Fourier transform.  In the present case, both $A$ and $B$ can  be
identified with $H^2(M_4,\ZZ)$ and the Fourier transform is a finite
sum:
\begin{equation}\label{tildpsi}\tilde\psi_b=C\sum_{a\in
H^2(M_4,\Z)}\exp(2\pi i (a,b))\psi_a.\end{equation} Here $C$ is a
constant and we write $\exp(2\pi i (a,b))$ for the perfect pairing
$H^2(M_4,\ZZ)\times H^2(M_4,\ZZ)\to U(1)$.

Let us interpret this formula in four-dimensional gauge theory.  In
$\Gad$ gauge theory on $M_4$, we can as before define a partition
function $Z_a$ by summing over bundles with a fixed characteristic
class $a\in H^2(M_4,\ZZ)$.  Identifying these with the $\psi_a$, we
find that under electric-magnetic duality the $Z_a$ must transform
by
\begin{equation}\label{ild}Z_b(-1/\tau)= C\sum_a\,\exp(2\pi i (b,a))
Z_a(\tau).\end{equation} We have incorporated the fact that (because
it exchanges the last two factors in $M_6=M_4\times S^1\times \tilde
S^1$) electric-magnetic duality inverts $\tau$, in addition to its
action on the label $a$.  This formula was first obtained in purely
four-dimensional terms in \cite{VW}, where more detail can be found.
Here we have given a six-dimensional context for this result.

If $G$ is a simply-laced and simply-connected Lie group, then its
GNO or Langlands dual group $G^\vee$ is precisely the adjoint group
$\Gad$. Apart from elementary constant factors that are considered
in \cite{VW}, the partition function of the theory with gauge group
$G$ is $Z_0$ (since $a$ must vanish if the gauge group is the
simply-connected form $G$), and the partition function of the theory
with gauge group $\Gad$ is $\sum_a Z_a$ (since all choices of $a$
are equally allowed if the gauge group is the adjoint form). As
noted in \cite{VW}, a special case of (\ref{ild}) is that $Z_0$
transforms under $\tau\to -1/\tau$ into a constant multiple of
$\sum_a Z_a$.  This assertion means that in this particular case the
$G$ and $G^\vee$ theories are dual. Other specializations of
(\ref{ild}) correspond to duality for forms intermediate between $G$
and $\Gad$, but in general (\ref{ild}) contains more information
than can be extracted from such special cases.

\remark The close analogy between the conformal blocks of the
six-dimensional $(0,2)$ model and those of the the level 1 WZW model
in two dimensions make one wonder if there might be an analog in six
dimensions of the WZW models at higher level. All one can say here
is that the usual $(0,2)$ model has appeared in string theory in
many ways and as of yet there is no sign of a hypothetical higher
level analog.

\subsection{What Is Next?}

In view of what we have said, if we specialize  to six-manifolds of
the form $M_6=M_4\times T^2$, where we keep the two-torus $T^2$
fixed and let only $M_4$ vary, the six-dimensional (0,2) theory
gives a good framework for understanding geometric Langlands.

We can do other things with this theory, since we are free to
consider more general six-manifolds.  This will be our topic in
Section 5.  But perhaps we should first address the following
question. Is this the end?  Or will physicists come back next year
and say that geometric Langlands should be derived from a theory
above six dimensions?

There is a precise sense in which six dimensions is the end.  It is
the maximum dimension for superconformal field theory, according to
an old result of Nahm \cite{N}.  To get farther, one needs a
different kind of theory.

If one wishes to go beyond six dimensions, the next stop is
presumably string theory (dimension ten).  Indeed, the existence and
most of the essential properties of the six-dimensional QFT that
underlies four-dimensional electric-magnetic duality are known
primarily from the multiple relations this theory has with string
theory.

\section{Geometric Langlands Duality For Surfaces}\label{surfdual}

\subsection{Circle Fibrations}

As we have discussed, one of the most basic properties of the
six-dimensional $(0,2)$ theory is that when formulated on
$M_6=M_5\times S^1$, it gives rise at long distances to
five-dimensional gauge theory on $M_5$.

The simplest generalization\footnote{The material in this section
was presented in more detail in lectures at the IAS in the spring of
2008. Notes  by D. Ben-Zvi can be found at
\url{http://www.math.utexas.edu/users/benzvi/GRASP/lectures/IASterm.html.}}
of this is to consider not a product $M_5\times S^1$, but a
fibration over $M_5$ with $S^1$ fibers:
\begin{equation}\label{zelf}\begin{matrix} S^1& \to & M_6\\
                                   &      & \downarrow\\
                                   &      &
                                   M_5.\end{matrix}\end{equation}
(For simplicity, we assume that the fibers are oriented.)  In this
situation, the long distance limit is still gauge theory on $M_5$,
with gauge group $G$.  But there is an important modification.

\def\CS{{\mathrm C} {\mathrm S}}
We pick on $M_6$ a Riemannian metric that is invariant under
rotation of the fibers of the $U(1)$ bundle $M_6\to M_5$.  Such a
metric determines a connection on this $U(1)$ bundle, and therefore
a curvature two-form $f\in \Omega^2(M_5)$.   Let $A$ be the gauge
field on $M_5$ (so $A$ is a connection on a $G$-bundle over $M_5$),
and let $\CS(A)$ be the associated Chern-Simons three-form. (As is
customary among physicists, we will normalize this form so that its
periods take values in $\Bbb{R}/2\pi\Bbb{Z}$.) Then the twisting of
the fibration $M_6\to M_5$ results in the presence in the long
distance effective action of an additional term $\Delta I$ that
roughly speaking is
\begin{equation}\label{omelf}\Delta I=\frac{i}{2\pi}\int_{M_5}f\wedge
\CS(A).\end{equation} To be more precise, one should define
$-i\Delta I$ as the integral of a certain Chern-Simons five-form for
the group $U(1)\times G$. This Chern-Simons five-form is associated
to an invariant cubic form on the Lie algebra of $U(1)\times G$ that
is linear on the first factor of this Lie algebra and quadratic on
the second.  Since $\Delta I$ is $i$ times the integral of a
Chern-Simons form, $\Delta I$ is well-defined and gauge-invariant
mod $2\pi i\Bbb{Z}$ assuming that $M_5$ is a compact manifold with
boundary. This ensures that $\exp(-\Delta I)$ is well-defined as a
complex number, so that it is possible to include a factor of
$\exp(-\Delta I)$ in the integrand of the path integral of
five-dimensional supersymmetric gauge theory on $M_5$. (Saying that
$\Delta I$ appears as a term in the effective action means precisely
that the integrand of the path integral has such a factor.)

\subsection{Allowing Singularities}\label{allowing}

However, it is natural to relax the conditions that we have imposed
so far. Describing $M_6$ as a $U(1)$ bundle over some base $M_5$
amounts to exhibiting a free action of the group $U(1)$ on $M_6$; if
such an action is given, one simply defines $M_5=M_6/U(1)$ and then
$M_6$ is a $U(1)$ bundle over $M_5$. Clearly, a more general
situation is to consider a six-manifold $M_6$ together with a
non-trivial action of the group $U(1)$. After possibly replacing
$U(1)$ by a finite quotient of itself (to eliminate a possible
finite subgroup that acts trivially), we can assume that $U(1)$ acts
freely on a dense open set in $M_6$.  The quotient $M_5=M_6/U(1)$ is
a five-manifold possibly with singularities where the $U(1)$ action
is non-free.

The above description, with the term (\ref{omelf}) in the effective
action, is applicable away from the non-free locus in $M_5$ (which
consists of the points in $M_5$ that correspond to non-free orbits
in $M_6$). Along the non-free locus, one should expect the gauge
theory description to require some kind of modification.  What sort
of modification is needed depends on how the $U(1)$ action fails to
be free.  $U(1)$ may have non-free orbits in codimension 2, 4, or 6,
and these non-free orbits may be either fixed points of the whole
group, or semi-free orbits whose stabilizer is a finite subgroup of
$U(1)$.  (To characterize the local behavior, one also needs to
specify the action of $U(1)$ in the normal space to the non-free
locus.  Further, though this will not be important for our purposes,
in general one wishes to allow the possibility of a $U(1)$ symmetry
that acts via a homomorphism to the $R$-symmetry group $Sp(4)$, in
addition to acting geometrically on $M_6$.)

\def\R{{\Bbb{R}}}
Thus,  for a full analysis of this problem, there are many
interesting cases to consider, most of which have not been analyzed
yet. A simple example is that $U(1)$ may act on $M_6$ with a fixed
point set of codimension 2, in which case $M_5$ is a manifold with
boundary. Thus a natural boundary condition in five-dimensional
supersymmetric gauge theory will have to appear.

For our purposes, we will consider just one situation, in which one
knows the appropriate modification of the effective field theory
that occurs near the exceptional set in $M_5$.
 This is the case of a codimension 4 fixed point locus $W$ such that the action
 of $U(1)$ on the normal space to $W$ can be
modeled by the natural action of $U(1)$ on $\CC^2\cong \Bbb{R}^4$.

Thus, focussing on the normal space to $W$, we take $U(1)$ to act on
$\CC^2$ by $(z_1,z_2)\to (e^{i\theta}z_1,e^{i\theta}z_2)$, for
$e^{i\theta}\in U(1)$. Clearly, this gives an action of $U(1)$ on
$\CC^2$ that is free except for an isolated fixed point at the
origin.  Somewhat less obvious -- but elementary to prove -- is that
the quotient $\CC^2/U(1)$ is actually a smooth manifold.  In fact,
it is a copy of $\Bbb{R}^3$:
\begin{equation}\label{zerk}\CC^2/U(1)\cong \Bbb{R}^3.\end{equation}

We can get this statement by taking a cone over the Hopf fibration.
The Hopf fibration is the $U(1)$ bundle $S^3\to S^2$.  A cone over
$S^3$ is $\R^4\cong \CC^2$, while a cone over $S^2$ is $\R^3$. So,
writing $0$ for the origin in $\R^4$ or $\R^3$,
$\R^4\backslash\{0\}$ is a $U(1)$ bundle over $\R^3\backslash\{0\}$.
Gluing back in the origin on both sides, we arrive at the assertion
(\ref{zerk}).

It follows from (\ref{zerk}) that if $U(1)$ acts on $M_6$ freely
except for a codimension 4 fixed point set $W$ as just described,
then $M_5=M_6/U(1)$ is actually a smooth manifold. A few simple
facts about the geometry of $M_5$ deserve attention.  One obvious
fact is that $W$ is naturally embedded as a codimension 3
submanifold of $M_5$.  Moreover, it is only away from $W$ that the
natural projection $M_6\to M_5=M_6/U(1)$ is a $U(1)$ fibration. This
projection thus gives a $U(1)$ bundle over $M_5\backslash W$, which
topologically cannot be extended over $M_5$.  The obstruction to
extending the $U(1)$ bundle can be measured as follows.  Let $S$ be
a small two-sphere in $M_5\backslash W$ that has linking number 1
with $W$.  (One can construct a suitable $S$ by choosing a normal
three-plane $N$ to $W$ at some chosen point $p\in W$ and letting $S$
consist of points in $N$ a distance $\epsilon$ from $p$, for some
small $\epsilon$.)  Then the $U(1)$ bundle over $M_5\backslash W$,
restricted to $S$, has first Chern class 1.

This fact can be expressed as an equation for the curvature two-form
$f$ of the $U(1)$ bundle over $M_5\backslash W$. As a form on
$M_5\backslash W$, $f$ is closed, obeying $\d f=0$. But $f$ has a
singularity along $W$ which can be characterized by the statement
\begin{equation}\label{helme}\d
{f}=2\pi\delta_W,\end{equation} where $\delta_W$ is the Poincar\'e
dual to $W$.

\subsubsection{Role Of $W$ In The Quantum Theory}

\def\L{{\mathcal L}}
In light of this information, let us consider now the $(0,2)$ theory
of type $G$ formulated on $M_6$, and its reduction to an effective
description on $M_5$.
 Away from $W$, as we have already discussed,
the effective theory on $M_5$ is simply supersymmetric gauge theory
with gauge group $G$, and with the additional interaction
(\ref{omelf}) that reflects the twisting of the fibration $M_6\to
M_5$. The gauge field is a connection on a $G$-bundle $E\to
M_5$.

However, there is an important and very interesting modification
along $W$.  This modification results from the fact that the
interaction $\Delta I$ is not well-defined in the usual sense. We
can define a Chern-Simons five-form on $M_5\backslash W$ for the
group $U(1)\times G$, but as $M_5\backslash W$ is not compact, the
integral of this form is not gauge-invariant, even modulo $2\pi $.

\def\A{{\mathcal A}}
\def\G{{\mathcal G}}
 Consequently, $\exp(-\Delta I)$, the
corresponding factor in the path integral, is not well-defined as a
complex number, but as a section of a certain complex line bundle
$\L$. $\L$ is a line bundle over the space of all $G$-valued gauge
fields, modulo gauge transformations, on $W$.  More exactly, $\L$ is
a line bundle over the space  of all connections on $E|_W$ modulo
gauge transformations ($E|_W$ is simply the restriction to $W$ of
the $G$ bundle  $E\to M_5$).  We write $\A$ for the space of
connections on $E|_W$ and $\G$ for the group of gauge
transformations; then $\L$ is a line bundle over the quotient
$\A/\G$ (or equivalently, a $\G$-invariant line bundle over $\A$).
In fact, $\L$ is the fundamental line bundle over $\A/\G$, often
loosely called the determinant line bundle. (The motivation for this
terminology is that if $G=SU(n)$ or $U(n)$ for some $n$, then $\L$
can be defined as the determinant line bundle of a $\bar\partial$
operator. It can also be defined as the Pfaffian line bundle of a
Dirac operator if $G=SO(n)$ or $Sp(2n)$.)

The characterization of $\L$ can be justified as follows.  The
interaction $\Delta I$ as defined in  (\ref{omelf}) does not depend
on a choice of gauge for the $U(1)$ bundle $M_6\backslash W\,\to\,
M_5\backslash W$, as it is written explicitly in terms of the $U(1)$
curvature $f$.  On the other hand, under an infinitesimal $G$ gauge
transformation $A\to A-\d_A\epsilon$, the Chern-Simons three-form
$\CS(A)$ transforms by $\CS(A)\to \CS(A)+\d X_2$, where $X_2$ is
known to physicists as the anomaly two-form (explicitly,
$X_2=(1/4\pi)\,\Tr\,\epsilon \d A$). Substituting this gauge
transformation law in (\ref{omelf}), integrating by parts, and using
(\ref{helme}), we see that under such a gauge transformation,
$\Delta I$ transforms by
\begin{equation}\label{zolf} \Delta I\to \Delta I-i\int_W
X_2,\end{equation} which is equivalent to saying that $\exp(-\Delta
I)$ should be understood as a section of the line bundle $\L$.

Physicists would describe this situation by saying that the factor
$\exp(-\Delta I)$ in the path integral has an anomaly under gauge
transformations that are non-trivial along $W$. The anomaly must be
canceled by incorporating in the theory another ingredient with an
equal and opposite anomaly. This additional ingredient must be
supported on $W$ (since away from $W$ we already know what is the
right effective field theory). The theory that does the job is the
two-dimensional (holomorphic) WZW model (or in other words, current
algebra, as explained in footnote \ref{able}) on $W$, at level 1.

This then is the secret of $W$: it supports this particular
two-dimensional quantum field theory.  This is the main fact that we
will use in interpreting recent mathematical results \cite{BF,Li,HN}
about instantons and geometric Langlands for surfaces.

\subsubsection{\it More Concrete Argument}\label{helpme} This
somewhat abstract argument can be replaced by a much more concrete
one if $G$ is a group of classical type, rather than an exceptional
group. (A similar analysis has been made independently for somewhat
related reasons in \cite{DHSV}.  See also \cite{Ch}. The following
discussion requires more detailed input from string theory than the
rest of the present article, and the reader may wish to jump to
section \ref{compac}.) The simplest case is that $G$ is $SU(n)$ or
even better $U(n)$. We use the fact \cite{Strominger} that the
$(0,2)$ model of $U(n)$ describes the low energy behavior of a
system of $n$ parallel M5-branes. We consider $M$-theory on
$\R^7\times {\mathrm{TN}}$, where $\mathrm{TN}$ is the Taub-NUT
space, a certain hyper-Kahler four-manifold that topologically is
$\R^4$. (It is described in detail in section \ref{tn}.)
$\mathrm{TN}$ has a $U(1)$ symmetry with $\mathrm{TN}/U(1)=\R^3$, as
suggested by eqn. (\ref{zerk}); we denote as $0$ the point in $\R^3$
that corresponds to the $U(1)$ fixed point in $\mathrm{TN}$. Inside
$\R^7\times\mathrm{TN} $, we consider $n$ M5-branes supported on
$\R^2\times\mathrm{TN}$ (for some choice of embedding
$\R^2\subset\R^7$); this gives a realization of the $(0,2)$ theory
of type $U(n)$ on $\R^2\times \mathrm{TN}$. We want to divide by the
$U(1)$ symmetry of $\mathrm{TN}$ to reduce the six-dimensional
$(0,2)$ model supported on the M5-branes to a five-dimensional
description. This may be done straightforwardly. For any
seven-manifold $Q_7$, $M$-theory on $Q_7\times\mathrm{TN}$ is
equivalent \cite{Townsend} to Type IIA superstring theory on
$Q_7\times \R^3$ with a D6-brane supported on $Q_7\times \{0\}$. So
$M$-theory on $\R^7\times \mathrm{TN}$ is equivalent to Type IIA on
$\R^7\times \R^3$ with a D6-brane supported at $\R^7\times \{0\}$.
In this reduction, the $n$ M5-branes on $\R^2\times \mathrm{TN}$
turn into $n$ D4-branes supported on $\R^2\times \R^3$. The low
energy theory on the D4-branes is $\mathcal N=4$ super Yang-Mills
theory with gauge group $U(n)$. The D4-branes intersect the D6-brane
on the Riemann surface $W=\R^2\times \{0\}$, and a standard
calculation (which uses the fact that the D4-branes and the D6-brane
intersect transversely on $W$) shows the appearance on $W$ of $U(n)$
current algebra at level 1. The behavior of the  $(0,2)$ model of
type $\mathrm{D}_n$ can be analyzed similarly by replacing $\R^7$ in
the starting point with $\R^5/\Z_2\times \R^2$.

\subsection{Compactification On A Hyper-Kahler
Manifold}\label{compac}

We are going to consider the $(0,2)$ theory in  a very special
situation.  We take $M_6=\R\times S^1\times X$, where $X$ will be a
hyper-Kahler four-manifold.  We think of $\R$ as parametrizing the
``time'' direction. On $M_6$, we take the obvious sort of product
metric, giving circumference $2\pi$ to $S^1$. We could take the
metric on $M_6$ to be of Euclidean signature (which would agree well
with some of our earlier formulas), but it is actually more elegant
in what follows to use a Lorentz signature metric, that is a metric
of signature $-+++++$, with  the negative eigenvalue corresponding
to the $\Bbb{R}$ direction.\footnote{One of the important general
facts about quantum field theory,
 as remarked in footnote
\ref{euclid}, is that in the world of unitary, physically sensible
quantum field theories with positive energy -- such as the
six-dimensional $(0,2)$ model considered here --  it is possible in
a natural way to formulate the ``same'' quantum field theory on a
space of Euclidean or Lorentzian signature. In the following
analysis, the main thing that we gain by using Lorentz signature is
that the supersymmetry generators are hermitian and the energy is
bounded below.}

The most obvious ordinary or bosonic conserved quantities in this
situation are the ones that act geometrically: the Hamiltonian $H$,
which generates translations in the $\R$ direction, the momentum
$P$, which generates rotations of $S^1$, and possible additional
conserved quantities associated with symmetries of $X$. The $(0,2)$
model also has a less obvious bosonic symmetry group; this is the
$R$-symmetry group $Sp(4)$, mentioned in Remark \ref{zorkox}.
Because $\R\times S^1$ is flat and $X$ is hyper-Kahler, so that
$M_6=\R\times S^1\times X$ admits covariantly constant spinor
fields, there are also unbroken supersymmetries. In fact, there are
eight unbroken supersymmetries $Q_\alpha$, $\alpha=1,\dots,8$; they
are hermitian operators that transform in the fundamental
representation of the $R$-symmetry $Sp(4)$ (which has real dimension
8). They commute with $H$ and $ P$, and obey a Clifford-like
algebra. With a suitable choice of normalizations and orientations,
this algebra reads
\begin{equation}\label{orsk}\{Q_\alpha,Q_\beta\}=2\delta_{\alpha\beta}\left(H-P\right).\end{equation}

\def\V{{\mathcal V}}
Accordingly, the operator $H-P$ is positive semi-definite; it can be
written in many different ways as the square of a Hermitian
operator.  States that are annihilated by $H-P$ are known as BPS
states and play a special role in the quantum theory \cite{WO}. We
write $\V$  for the space of BPS states. $\V$ admits an action of
$U(1)\times Sp(4)$ (or possibly a central extension thereof), where
$U(1)$ is the group of rotations of $S^1$ and $Sp(4)$ is the
$R$-symmetry group. The center of $Sp(4)$ is generated by an element
of order 2 that we denote as $(-1)^F$; it acts as $+1$ or $-1$ on
bosonic or fermionic states, respectively. So in particular, $\V$ is
$\Z_2$-graded by the eigenvalue of $(-1)^F$.  We refer to $\V$, with
its action of $U(1)\times Sp(4)$, as the spectrum of BPS states. One
important general fact is that $P$ is bounded below as an operator
on $\V$; indeed, on general grounds, $H$ is bounded below in the
full Hilbert space of the $(0,2)$ theory, while $H=P$ when
restricted to $\V$.

Certain features of the spectrum of BPS states are ``topological
invariants,'' that is, invariant under continuous deformations of
parameters. (In the present problem, the relevant parameters are the
moduli of the hyper-Kahler metric of $X$.)  The most obvious such
invariant is the ``elliptic genus,'' $F(q)=\Tr_{\V}\,(-1)^Fq^P$,
where $q$ is a complex number with $|q|<1$.  (It has modular
properties, since it can be represented by the partition function of
the $(0,2)$ model on $\Sigma\times X$, where $\Sigma$ is an elliptic
curve whose modular parameter is $\tau=\ln q/2\pi i$.)  $F(q)$ is
invariant under smooth deformation of the spectrum by virtue of the
same arguments that are usually used to show that the index of a
Fredholm operator is invariant under deformation.

In the present problem, the whole spectrum of BPS states, and not
only the index, is invariant under deformation of the hyper-Kahler
metric of $X$.  One approach to proving this uses the fact that $\V$
can be characterized as the cohomology of $\mathcal Q$, where
$\mathcal Q$ is any complex linear combination of the hermitian
operators $Q_\alpha$ that squares to zero. Picking any one complex
structure on $X$ (from among the complex structures that make up the
hyper-Kahler structure of $X$), one makes a judicious choice of
$\mathcal Q$ to show that the spectrum of BPS states is invariant
under deformations of the Kahler metric of $X$ (keeping the chosen
complex structure fixed). Repeated moves of this kind (specializing
at each stage to a different complex structure and therefore a
different choice of $\mathcal Q$) can bring about arbitrary changes
of the hyper-Kahler metric of $X$, so the spectrum of BPS states is
independent of the moduli of $X$.

In our discussion in section \ref{tn}, we compare two computations
of $\V$ in two different regions of the moduli space of hyper-Kahler
metrics on $X$.  The results must be equivalent in view of what has
just been described.

\remark There is some sleight of hand here, as the arguments above
have assumed $X$ to be compact, and we will use the results for
noncompact $X$.  So some refinement of the arguments is actually
needed.

\subsection{Taub-NUT Spaces}\label{tn}

Now the question arises of what sort of hyper-Kahler four-manifold
$X$ we will select in the above construction.

We will choose $X$ to admit a triholomorphic $U(1)$ symmetry, that
is, a $U(1)$ symmetry that preserves the hyper-Kahler structure of
$X$.  (Among other things, this ensures that this $U(1)$ also
commutes with the unbroken supersymmetries $Q_\alpha$ of eqn.
(\ref{orsk}).)  Hyper-Kahler four-manifolds with triholomorphic
$U(1)$ symmetry are highly constrained \cite{Lindstrom:1983rt}. The
general form of the metric is \begin{equation}\label{genf} \d
s^2=U\,\d\vec x\cdot \d\vec x +\frac{1}{U}(\d\theta+\vec \omega\cdot
\d\vec x)^2,\end{equation} where $\vec x$ parametrizes $\R^3$, $U$
is a harmonic function on $\R^3$, and (away from singularities of
$U$) $\theta$ is an angular variable that parametrizes the $U(1)$
orbits.

\def\N{{\mathcal N}}
This form of the metric shows that the quotient space $X/U(1)$
(assuming $X$ is complete) is equal to $\R^3$.  Indeed, the natural
projection $X\to X/U(1)$, which was considered in section
\ref{allowing}, has a special interpretation in this situation.
  It is
the hyper-Kahler moment map $\vec\mu$ and it is a surjective map to
$\R^3$:
\begin{equation}\label{city}\vec\mu:X\to\R^3.\end{equation}

The most obvious hyper-Kahler four-manifold with a triholomorphic
$U(1)$ symmetry is $\R^4$.  This corresponds to the choice
$U=1/2|\vec x|$.  The $U(1)$ action on $\R^4$ has a fixed point at
the origin (where $U$ has a pole and the radius of the $U(1)$ orbits
vanishes, according to (\ref{genf})).  This fixed point is precisely
of the sort considered in section \ref{allowing}.  To verify this,
begin with the fact that the rotation group of $\R^4$ has
$SU(2)_L\times SU(2)_R$ for a double cover; $SU(2)_L$ and $SU(2)_R$
are two copies of $SU(2)$. We can pick a hyper-Kahler structure on
$X$ compatible with its flat metric such that  $SU(2)_L$ rotates the
three complex structures and $SU(2)_R$ preserves them. We simply
take $U(1)$ to be a subgroup of $SU(2)_R$. Then, upon picking a
complex structure on $\R^4$ that is invariant under $SU(2)_L\times
U(1)$ (this complex structure
 is {\it not} part of its $U(1)$-invariant
hyper-Kahler structure), we can identify $\R^4$ with $\Bbb{C}^2$ and
$U(1)$ acts in the natural way $(z_1,z_2)\to
(e^{i\theta}z_1,e^{i\theta}z_2)$. This then is the situation that
was considered in section \ref{allowing}, and the statement
(\ref{city}) gives a hyper-Kahler perspective on the fact that the
quotient $\R^4/U(1)$ is $\R^3$, as was asserted in (\ref{zerk}).

\def\CC{{\Bbb{C}}}
\def\TN{{\mathrm{TN}}}
Although $\R^4$ has the properties we need from a topological point
of view, there is a different hyper-Kahler metric on $\R^4$  that
will be more useful for our application in section \ref{twoways}.
This is the Taub-NUT space, which we will call $\TN$. To describe
$\TN$ explicitly, we simply choose $U$ to be
\begin{equation}U=\frac{1}{R^2}+\frac{1}{2|\vec x|},\end{equation}
where $R$ is a constant.   Looking at (\ref{genf}), the
interpretation of $R$ is easy to understand: the $U(1)$ orbits have
circumference $2\pi/\sqrt U$, which at infinity approaches $2\pi R$.
The flat metric on $\R^4$ is recovered in the limit $R\to\infty$; in
$\R^4$, of course, the circumference of an orbit diverges at
infinity.

Accordingly, the hyper-Kahler metric on $\TN$ is quite different at
infinity from the usual flat hyper-Kahler metric on $\R^4$. However,
in one sense the difference is  subtle.  If we pick any one of the
complex structures that make up the hyper-Kahler structure, then it
can be shown that, as a complex symplectic manifold in this complex
structure, $\TN$ is equivalent to $\R^4\cong \CC^2$.

A more general choice of $X$ is also important.  First of all,
naively we could pick an integer $k>1$ and take $X=\R^4/\Z_k$, where
$\Z_k$ is the subgroup of $U(1)$ consisting of points of order $k$.
Certainly $\R^4/\Z_k$ has a (singular) hyper-Kahler metric with a
triholomorphic $U(1)$ symmetry.  The singularity at the origin of
$\R^4/\Z_k$ is known as an ${\mathrm A}_{k-1}$ singularity. It is
possible to make a hyper-Kahler resolution of this singularity,
still with a triholomorphic $U(1)$ symmetry.  This is accomplished
by picking $k$ points $\vec x_1,\dots\vec x_k$ in $\R^3$ and setting
$U=\frac{1}{2}\sum_{j=1}^k 1/|\vec x-\vec x_j|$.  This gives a
complete hyper-Kahler manifold which is smooth if the $\vec x_j$ are
distinct.  As a complex symplectic manifold in one complex
structure, it can be described by an equation
\begin{equation}\label{zort} uv=f(w),\end{equation}
where $f(w)$ is a $k^{th}$ order monic polynomial.  This is the
usual complex resolution of the ${\mathrm A}_{k-1}$ singularity. In
this description, the holomorphic symplectic form is $\d u \wedge \d
v/f'(w)$, and the triholomorphic $U(1)$ symmetry is $u\to \lambda
u$, $v\to\lambda^{-1} v$.

However, again, a generalization is more convenient for our
application in section \ref{twoways}.  We simply add a constant to
$U$ and take
\begin{equation}\label{tronk}U=\frac{1}{R^2}+\frac{1}{2}\sum_{j=1}^k\frac{1}{|\vec
x-\vec x_j|}.\end{equation}  This gives a complete hyper-Kahler
manifold, originally constructed in \cite{Hawking:1976jb}, that we
call the multi-Taub-NUT space and denote as $\TN_k$.

As a complex symplectic manifold in any one complex structure,
$\TN_k$ is independent of the parameter $R$ and coincides with the
usual resolution (\ref{zort}) of the ${\mathrm A}_{k-1}$
singularity.  However, the addition of a constant to $U$ markedly
changes the behavior of the hyper-Kahler metric at infinity.   Just
as in the $k=1$ case that was considered earlier, the asymptotic
value at infinity of the circumference of the fibers of the
fibration $\TN_k\to \R^3$ is  $2\pi R$.

The space $\TN_k$ is smooth as long as the $\vec x_j$ are distinct.
When $r$ of them coincide, an ${\mathrm A}_{r-1}$ singularity
develops, that is, an orbifold singularity of type $\R^4/\Z_r$.

In general, for $\vec x\to\vec x_j$, we have $U\to\infty$.  So at
those points, and only there, the radius of the $U(1)$ fibers
vanishes. The $k$ points $\vec x=\vec x_j$ are, accordingly, the
fixed points of the triholomorphic $U(1)$ action.

\subsubsection{A Note On The Second Cohomology}\label{note}

We conclude this subsection with some technical remarks that will be
useful in section \ref{twoways} (but which the reader may choose to
omit). Topologically, $\TN_k$ is, as we have noted, the same as the
resolution of the ${\mathrm A}_{k-1}$ singularity. A classic result
therefore identifies $H^2(\TN_k,\Z)$ with the root lattice of the
group ${\mathrm A}_{k-1}=SU(k)$.

\def\L{\mathcal L}
However, $\TN_k$ is not compact and one should take care with what
sort of cohomology one wants to use.  It turns out that  another
natural definition is useful.  We define an abelian group $\Gamma_k$
as follows: an element of $\Gamma_k$ is a unitary line bundle
${\mathcal L}\to \TN_k$ with anti-selfdual and square-integrable
curvature and whose connection has trivial holonomy when restricted
to a fiber at infinity of $\vec\mu:\TN_k\to \R^3$. $\Gamma$ is a
discrete abelian group with a natural and integer-valued quadratic
form, defined as follows; if ${\mathcal L}$ is a line bundle with
anti-selfdual curvature $F$, we define $(\L,\L)=-\int_{\TN_k}F\wedge
F/4\pi^2$.

It turns out that $\Gamma_k\cong \Z^k$, with the quadratic form
corresponding to the quadratic function of $k$ variables
$y_1^2+y_2^2+\dots+y_k^2$. (A basis of $\Gamma_k$ is described in
section \ref{basis}.) Thus, $\Gamma$ corresponds to the weight
lattice of the group $U(k)$.  In many string theory problems
involving Taub-NUT spaces, one must use $\Gamma_k$ as a substitute
for $H^2(\TN_k,\Z)$, which does not properly take into account the
behavior at infinity.

This is notably true if one considers the $(0,2)$ model on
$M_6=W\times \TN_k$, for $W$ a Riemann surface. In section
\ref{confblock}, we explained that the $(0,2)$ model on a compact
six-manifold $M_6$ has a space of conformal blocks that is obtained
by quantizing, in a certain sense, the finite abelian group
$H^3(M_6,\ZZ)$.  For $M_6=W\times \TN_k$, the appropriate substitute
for this group is
\begin{equation}\label{gurer}\tilde H^3(W\times
\TN_k,\ZZ)=H^1(W,\ZZ)\otimes\Gamma_k.\end{equation}

\subsubsection{Basis Of $\Gamma_k$}\label{basis}
 It is furthermore true that $\Gamma_k$ has a natural
basis corresponding to the $U(1)$ fixed points $\vec x_j$,
$j=1,\dots,k$. To show this, we first describe a dual basis of
noncompact two-cycles.  For $j=1,\dots,k$, we let $\ell_j$ be a path
in $\R^3$ from $\vec x_j$ to $\infty$, not passing through any $\vec
x_r$ for $r\not=j$.  Then we set $C_j=\vec\mu^{-1}(\ell_j)$. $C_j$
is a noncompact two-cycle that is topologically $\R^2$.  A line
bundle $\L$ that represents a point in $\Gamma_k$ is trivialized at
infinity on $C_j$ because its connection is trivial on the fibers of
$\vec\mu$ at infinity.  So we can define an integer
$\int_{C_j}c_1(\L)$.  One can pick a basis of $\Gamma_k$ consisting
of line bundles $\L_r$ such that $\int_{C_j}c_1(\L_r)=\delta_{jr}$.
(The $\L_r$ are described explicitly in \cite{Wit1}.)

Now let us reconsider the definition of $\tilde H^3(W\times
\TN_k,\ZZ)$ in (\ref{gurer}).   From what we have just said,
$H^1(W,\ZZ)\otimes\Gamma_k$ has a natural decomposition as the
direct sum of copies $H^1_{(j)}(W,\ZZ)$ of $H^1(W,\ZZ)$ associated
with the fixed points:
\begin{equation}\tilde H^3(W\times\TN_k)=\oplus_{j=1}^k
H^1_{(j)}(W,\ZZ).\end{equation} Upon quantization, this means that
the space of conformal blocks of the $(0,2)$ model on $W\times
\TN_k$ is the tensor product of $k$ factors, each of them isomorphic
to the space of conformal blocks in the level 1 WZW model
(associated with the group $G$) on $W$.  The factors are naturally
associated to the $U(1)$ fixed points.

\remark\label{itelater}  Similarly, we can enrich the definition of
the two-dimensional characteristic class $a$ of a $\Gad$ bundle over
$\TN_k$.  Normally, $a$ takes values in $H^2(\TN_k,\ZZ)$. However,
suppose $E\to \TN_k$ is a $\Gad$ bundle that is trivialized over
each fiber at infinity of $\vec\mu:\TN_k\to\R^3$.  Then $E$ is
trivialized at infinity on each $C_j$, so one can define a pairing
$a_j=\langle a,C_j\rangle$ for each $j$; the $a_j$ take values in
$\ZZ$.  Equivalently, we can consider $a$ as an element of $\tilde
H^2(\TN_k,\ZZ)=\Gamma_k\otimes\,_\Z\ZZ$. This also has an analog if
we are given a conjugacy class ${\mathcal C}\subset \Gad$ and the
monodromy of $E$ on each fiber at infinity lies in $\mathcal C$.
Then one can define a ${\mathcal C}$-dependent torsor for the group
$\tilde H^2(\TN_k,\ZZ)$, and one can regard $a$ as taking values in
this torsor.  Concretely, this means that, once we pick a path in
$\Gad$ from $\mathcal C$ to the identity (an operation that
trivializes the torsor), we can define the elements $a_j\in \ZZ$ as
before. Two different paths from $\mathcal C$ to the identity would
differ by a closed loop in $\Gad$, corresponding to an element $b\in
\ZZ$; if we change the trivialization of the torsor by changing the
path by $b$, then the $a_j$ are shifted to $a_j+b$.  ($b$ is the
same for all $j$, since the regions at infinity in the two-cycles
$C_j$ can be identified, by taking the paths $\ell_j$ to coincide at
infinity.)

\subsection{Two Ways To Compute The Space Of BPS
States}\label{twoways}

Now we are going to study in two different ways the space of BPS
states of the $(0,2)$ model formulated on
\begin{equation}\label{jury}M_6=\R\times S^1\times \TN_k.\end{equation}
The results will automatically be equivalent, as explained at the
end of section \ref{compac}.

   $M_6$ admits an action of $U(1)\times U(1)'$ (the product of two
factors of $U(1)$), where $U(1)$ acts by rotation of $S^1$, and
$U(1)'$ is the triholomorphic symmetry of $\TN_k$.   We choose a
product metric on $M_6$, such that $S^1$ has circumference $2\pi S$,
and $\TN_k$ has a hyper-Kahler metric in which the $U(1)'$ orbit has
asymptotic circumference $2\pi R$. In section \ref{compac}, we took
$S=1$; in any event, because the $(0,2)$ model is conformally
invariant, only the ratio $R/S$ is relevant.

$U(1)$ and $U(1)'$ play very different roles in the formalism
because of the structure of the unbroken supersymmetry algebra,
which we repeat for convenience:
\begin{equation}\label{borsk}\{Q_\alpha,Q_\beta\}=2\delta_{\alpha\beta}\left(H-P\right).\end{equation}
Here $P$ is the generator of the $U(1)$ symmetry.  It appears in the
definition of the elliptic genus $F(q)={\mathrm {Tr}}_\V
\,q^P(-1)^F$, where $\V$ is the space of BPS states.  The function
$F(q)$ has modular properties, so if it is nonzero (as will turn out
to be the case), there are BPS states with arbitrarily large
eigenvalues of $P$. By contrast, it turns out that $U(1)'$ acts
trivially on $\V$.

One of our two descriptions of $\V$ will be good for $S\to 0$ or
equivalently $R\to\infty$; the other description will be good for
$R\to 0$ or equivalently $S\to\infty$. Comparing them will give a
new perspective on the results of \cite{BF,Li,HN}.

\subsubsection{Description I}\label{firstone}

For $S\to 0$, the low energy description is by gauge theory on
$M_6/U(1)=\R\times \TN_k$.  As $U(1)$ acts freely, we need not be
concerned here with the behavior at fixed points.  As the metric of
$M_6$ is a simple product $S^1\times M_5$ (with $M_5=\R\times
\TN_k=M_6/U(1)$), we also need not worry about the interaction
described in eqn. (\ref{omelf}).  So we simply get maximally
supersymmetric Yang-Mills theory on $\R\times \TN_k$, with gauge
group $G$.

In formulating gauge theory on $\R\times \TN_k$, we specify up to
conjugacy the holonomy $U$ of the gauge field over a fiber at
infinity of $\vec\mu:\TN_k\to\R^3$.   This choice (which has a
six-dimensional interpretation)
 leads to an important bigrading of the physical Hilbert space
$\mathcal H$ of the theory and in particular of the space $\mathcal
V$ of BPS states.  First of all, let $H$ be the subgroup of $G$ that
commutes with $U$. Classically, one can make a gauge transformation
that approaches at infinity a constant element of $H$; quantum
mechanically, to avoid infrared problems, the constant should lie in
the center of $H$.  So the center of $H$ acts on $\mathcal H$ and
$\mathcal V$.  We call this the electric grading.  (The center of
$H$ is, of course, abelian, and the eigenvalues of its generators
are called electric charges.)

A second ``magnetic'' grading arises for topological reasons.
 When $U\not=1$, the
topological classification of finite energy gauge fields on $\TN_k$
becomes more elaborate.  Near infinity on $\TN_k$, the monodromy
around $S^1$ reduces the structure group from $G$ to $H$, and the
bundle can be pulled back from an $H$-bundle over the region near
infinity on $\R^3$. Infinity on $\R^3$ is homotopic to $S^2$, so we
get an $H$-bundle over $S^2$. The Hilbert space of the theory is
then graded by the topological type of the $H$-bundle.  We call this
the magnetic grading. (Its components corresponding to $U(1)$
subgroups of $H$ are called magnetic charges.)

According to \cite{GNO},  electric-magnetic duality exchanges the
electric and magnetic gradings.  In our context, this will mean that
the electric grading in Description I matches the magnetic grading
in Description II, and viceversa. In the simplest situation, if $U$
is generic, then $H$ is a maximal torus $T$ of $G$; the electric and
magnetic gradings correspond to an action of $T$ and $T^\vee$,
respectively.

Actually, to extract the maximum amount of information from the
theory, we want to allow an arbitrarily specified value of the
two-dimensional characteristic class $a$.
  As described in Remark
\ref{itelater}, $a$ takes values in a certain torsor for $\tilde
H^2(\TN_k,\ZZ)$, which means, modulo a trivialization of the torsor,
that $a$ assigns an element of $\ZZ$ to each fixed point. (The
origin of $a$ in six dimensions was discussed in section
\ref{confblock}.) Roughly speaking, allowing arbitrary $a$  means
that we do $\Gad$ gauge theory, but there is a small twist: to
extract the most information, we divide by only those gauge
transformations that can be lifted to the simply-connected form $G$.
This means that the monodromy $U$ can be regarded as an element of
$G$ (up to conjugacy), and similarly that in Description II, we meet
representations of the Kac-Moody group of $G$ (not $\Gad$).

In gauge theory on $\R\times \TN_k$, $U(1)'$ acts geometrically,
generating the triholomorphic symmetry of $\TN_k$.  But how does
$U(1)$ act? The answer to this question is that in this description,
the generator $P$ of $U(1)$ is equal to the instanton number $I$.
(This fact is deduced using string theory.) The instanton number is
defined via a familiar curvature integral, normalized so that on a
compact four-manifold and with a simply-connected gauge group, it
takes integer values.  In the present context, the values of the
instanton number are not necessarily integers, because $\TN_k$ is
not compact.  The analog of integrality in this situation is the
following. First, one should add to the instanton number $I$ a
certain linear combination of the magnetic charges (with
coefficients given by the logarithms of the monodromies).  Let us
call the sum $\widehat I$. Then there is a fixed real number $r$,
depending only on the monodromy at infinity and the characteristic
class $a$, such that $\widehat I$ takes values in $r+\Bbb Z$.
  So in this description, eigenvalues of $P$ are not
necessary integers, but (for bundles with a fixed $a$ and $U$) a
certain linear combination of the eigenvalues of $P$ and the
magnetic charges are congruent to each other modulo integers.

Since $P'$ generates the $U(1)'$ symmetry of $\TN_k$, one might
expect its eigenvalues to be integers, but here we run into the
electric charges.   There is an operator that generates the
triholomorphic symmetry and whose eigenvalues are integers; it is
not simply $P'$ but the sum of $P'$ and a central generator of $H$
(this generator is the logarithm of the monodromy at infinity), or
in other words the sum of $P'$ and a linear combination of electric
charges.

What are BPS states in this description?  Classically, the minimum
energy fields of given instanton number are the instantons -- that
is the gauge fields that are independent of time and are
anti-selfdual connections on $\TN_k$. Instantons on $\TN_k$ have
recently been studied by $D$-brane methods
\cite{Cherkis1,Cherkis2,Wit1}.  In particular \cite{Cherkis1},
certain components of the moduli space $\M$ of instantons on
$\TN_k$, when regarded as complex symplectic manifolds in one
complex structure, coincide with components of the moduli space of
instantons on the corresponding ALE space (the resolution of
$\R^4/\Z_k$). All components of instanton moduli space on the ALE
space arise in this way, but there are also components of instanton
moduli space on $\TN_k$ that have no analogs for the ALE space.
(According to \cite{Cherkis1} and as explained to me by the author
of that paper, these are the components of nonzero magnetic charge,
corresponding to nonzero electric charge in Description II.)

\def\LL{{\mathrm {L}}}
An instanton is a classical BPS configuration, but to construct
quantum BPS states, we must, roughly speaking, take the cohomology
of the instanton moduli space $\M$.  Actually, $\M$ is not compact
and by ``cohomology,'' we mean in this context the space of $\LL^2$
harmonic forms on $\M$.  (These are relevant for essentially the
same reasons that they entered in one of the early tests of
electric-magnetic duality \cite{Sen}.)  So $\V$ is the space of
$\LL^2$ harmonic forms on $\M$.   Of course, to construct $\V$ we
have to include contributions from all components $\M_n$ of $\M$:
\begin{equation}\label{zomo}\V=\oplus_n H^*_{\LL^2\,\,{\mathrm
{harm}}}(\M_n),\end{equation} where we write $H^*_{\LL^2\,\,{\mathrm
{harm}}}$ for the space of $\LL^2$ harmonic forms. The action of $P$
on $\V$ is multiplication by the instanton number, and similarly the
magnetic grading is determined by the topological invariants of the
bundles parametrized by a given $\M_n$.  $P'$ and the electric
charges act trivially on $\V$ because they correspond to continuous
symmetries of $\M_n$ that act trivially on its cohomology.

Each $\M_n$ is a hyper-Kahler manifold, and accordingly the group
$Sp(4)$ -- which in the present context is the $R$-symmetry group
(as explained in Remark \ref{zork}) -- acts on the space of $\LL^2$
harmonic forms on $\M_n$ and hence on $\V$. However, as in similar
problems \cite{Hitchin}, it seems likely that $Sp(4)$ acts trivially
on these spaces. (This is equivalent to saying that $\LL^2$ harmonic
forms exist only in the middle dimension and are of type $(p,p)$ for
every complex structure.)  This would agree with what one sees on
the other side of the duality, which we consider next.

\remark If we simply replace $\TN_k$ by $\R^4$ (with its usual
metric) in this analysis, we learn in the same way that  BPS states
of the $(0,2)$ model on $\R\times S^1\times \R^4$ correspond to
$\LL^2$ harmonic forms on instanton moduli space on $\R^4$, with its
usual metric. The same holds with an ALE space instead of $\R^4$.
The advantage of $\TN_k$ over $\R^4$ or an ALE space is that there
is an alternative second description.
\bigskip

\subsubsection{Description II}\label{secondone}

The other option is to take $R\to 0$.  In this case, the fibers of
$\vec\mu:\TN_k\to\R^3$ collapse, so to go over to a gauge theory
description, we replace $\TN_k$ by $\R^3$, with special behavior at
the $U(1)$ fixed points $\vec x_j$, $j=1,\dots,k$, where holomorphic
WZW models will appear.  We get a second description, then, in terms
of maximally supersymmetric gauge theory on $M_5=\R\times S^1\times
\R^3$, with level 1 holomorphic WZW models of type $G$ supported on
the $k$ two-manifolds $W_j=\R\times S^1\times \vec x_j$,
$j=1,\dots,k$.

Once again, we must specify the holonomy $U$ at infinity of the
gauge field around $S^1$.  This is simply the same as the
corresponding holonomy at infinity in Description I.  Suppose for a
moment that $U$ is trivial. Then we also must pick, for each $\vec
x_j$, $j=1,\dots,k$, an integrable representation of the affine
Kac-Moody algebra of $G$ at level 1. For a simply-laced and
simply-connected group, the integrable representations are
classified by characters of the center $\ZZ$ of $G$, or, as there is
a perfect pairing $\ZZ\times\ZZ\to U(1)$, simply by $\ZZ$. So for
each $j$, we must give an element $a_j\in\ZZ$.  This is precisely
the data that we obtained in Description I from the characteristic
class $a\in \tilde H^2(\TN_k,\ZZ)$.  Since the second homology group
of $\R\times S^1\times \R^3$ vanishes, there is no two-dimensional
characteristic class to be chosen in Description II (matching the
fact that there was no Kac-Moody representation in Description I).

More generally, for any $U$, we can canonically pick up to
isomorphism a $G$-bundle on $\R\times S^1\times \R^3$ with that
monodromy at infinity, namely a flat bundle with holonomy $U$ around
$S^1$.  In the presence of this flat bundle, the Kac-Moody algebra
on each $S^1\times\vec x_j$ is twisted; if $\theta$ is an angular
parameter on $S^1$, then instead of the currents obeying
$J(\theta+2\pi)=J(\theta)$, they obey
$J(\theta+2\pi)=UJ(\theta)U^{-1}$.  The representations of this
twisted Kac-Moody algebra at level 1 are a torsor for $\ZZ$ -- the
same torsor that we met in Remark \ref{itelater}. The torsor is the
same for each $j$ since each Kac-Moody algebra is twisted by the
same $U$. (The torsor property means concretely that the
representations of the Kac-Moody algebra are permuted if $U$
undergoes monodromy around a noncontractible loop in $G_{\mathrm
ad}$.)

In Description II, $P$ generates the rotations of $S^1$.  For
reasons that will become apparent, what is important is how $P$ acts
on the representations of the Kac-Moody algebra. In the Kac-Moody
algebra, $P$ corresponds to the operator -- usually called $L_0$ --
that generates a rotation of the circle. First set $U=1$. Then $L_0$
has integer eigenvalues in the vacuum representation of the
Kac-Moody algebra (that is, the representation whose highest weight
is $G$-invariant). In a more general representation (but still at
$U=1$), $L_0$ has eigenvalues that are congruent mod $\Z$ to a fixed
constant $r$ that depends only on the highest weight.  This matches
the fact that, in Description  I (at $U=1$) the instanton number
takes values in $r+\Bbb Z$ where $r$ depends only on the
characteristic class $a$. In the Kac-Moody theory, when the twisting
parameter $U$ is varied away from 1, the eigenvalues of $L_0$ shift.
However (recalling that $H$ is the commutant of $U$ in $G$), one can
add to $L_0$ a linear combination of the generators of $H$  to make
an operator $\widehat L_0$ with the property that in a given
representation of the twisted Kac-Moody algebra, its eigenvalues are
congruent mod $\Bbb Z$. Thus, electric charges play precisely the
role in Description II that magnetic charges play in Description I.

On the other hand, in Description II, $P'$ is the instanton number
of a $G$-bundle on the initial value surface $S^1\times\R^3$.  If
the monodromy $U$ at infinity is trivial, then $P'$ is
integer-valued, just as in Description I.  In general, for any $U$,
a certain linear combination of $P'$ and the magnetic charges (with
coefficients given as usual by the logarithms of the monodromies)
takes integer values. This mirrors the fact that in Description I, a
linear combination of $P'$ and the electric charges takes integer
values.

What is the space of BPS states in Description II?  Supported on
$\R\times S^1\times \vec x_j$ for each $j=1,\dots,k$, there is a
level 1 Kac-Moody module ${\cal W}_j$. This module has $H=P$ for all
states (mathematically, the representation theory of affine
Kac-Moody algebras is usually developed with a single $L_0$
operator, not two of them), and consists entirely of BPS states. The
space of BPS states is simply $\V=\otimes_{j=1}^k{\mathcal W}_j$. In
particular, as the $R$-symmetry group $Sp(4)$ acts trivially on the
${\mathcal W}_j$, it acts trivially on $\V$.  The analogous
statement in Description I was explained at the end of section
\ref{firstone}.

Comparing  the results of the two descriptions, we learn
that
\begin{equation}\otimes_{j=1}^k{\mathcal W}_j=\oplus_n
H^*_{\LL^2\,\,{\mathrm {harm}}}(\M_n).\end{equation} The right hand
side is graded by instanton number and magnetic charge, and the left
hand side by $L_0$ and electric charge. This equivalence closely
parallels a central claim in \cite{BF,Li,HN}.

A couple of differences may be worthy of note. Our description uses
$\LL^2$ harmonic forms; different versions of cohomology are used in
recent mathematical papers. Also, our instantons live on $\TN_k$
with its hyper-Kahler metric, not on the resolution of the ${\mathrm
A}_{k-1}$ singularity.  This does not affect $\M_n$ as a complex
symplectic manifold (as long as one considers on the $\TN_k$ side
judiciously chosen components of the moduli space \cite{Cherkis1}),
but it certainly affects the hyper-Kahler metric of $\M_n$ and
therefore the condition for an $\LL^2$ harmonic form.  The
components $\M_n$ of instanton moduli space on $\TN_k$ that do not
have analogs on the resolution of the ${\mathrm A}_{k-1}$
singularity are also presumably important.

One may wonder why we do not get additional BPS states from
quantizing the moduli space of instantons, as we did in Description
I.  This can be understood as follows.

 Generically, curvature breaks all supersymmetry. In Description I,
because the curvature of $\TN_k$ is anti-selfdual, it leaves
unbroken half of the supersymmetry. The half that survives is
precisely the supersymmetry that is preserved by an instanton (since
an instanton bundle also has anti-selfdual curvature).  Hence
instantons are supersymmetric and must be considered in constructing
the space of BPS states.  By contrast, in Description II, there is
no curvature to break supersymmetry.  Instead, there is a coupling
(\ref{omelf}), which (when extended to include fields and terms that
we have omitted) leaves unbroken half the supersymmetry, but a
different half from what is left unbroken by anti-selfdual
curvature.  The result is that in Description II, instantons are not
supersymmetric.

So in Description II, the instanton number and similarly the
magnetic charges annihilate any BPS state.   This implies that $P'$
and the magnetic charges annihilate $\mathcal V$ in Description II,
just as $P'$ and the electric charges do in Description I.

\subsubsection{A Note On The Dual Group}\label{lowdown}  The reader
may be puzzled by the fact that in this analysis of two ways to
describe the space of BPS states, we have not mentioned the dual
group $G^\vee$.  The reason for this is that for simplicity, we have
limited ourselves to the case that $G$ is simply-laced.  When this
is so, $G$ and $G^\vee$ have the same Lie algebra.  Instead of
merely comparing $G$ and $G^\vee$ theories, we can learn more, as
explained in section \ref{confblock}, by considering $\Gad$ bundles
with an arbitrary two-dimensional characteristic class $a$.  This is
what we have done.

For groups that are not simply-laced, a slight variant of the
construction that we have used is available \cite{Vafa}.  The basic
idea is that outer automorphisms of a simply-laced group $G$ can
appear as symmetries of the $(0,2)$ model of type $G$ in six
dimensions. To a simple but not simply-laced Lie group $H$, one
associates a pair $(G,\lambda)$, where $G$ is simply-laced and
$\lambda$ is an outer automorphism of $G$. The $(0,2)$ model of type
$G$, when formulated on $M_6=M_5\times S^1$, can be ``twisted'' by
$\lambda$ in going around $S^1$, in which case the low energy limit
on $M_5$ is maximally supersymmetric gauge theory of type $H^\vee$.
Now consider $M_6=M_4\times S^1\times \tilde S^1$, with circles of
radius $S$ and $R$  and a twist by $\lambda$ around $S^1$. Repeating
the analysis of section \ref{nonca}, we get at long distances on
$M_4\times \tilde S^1$ a description by $H^\vee$ gauge theory; this
further reduces on $M_4$ to a description by $H^\vee$ gauge theory
with coupling parameter $\tau^\vee=iS/R$.   Alternatively, we get a
description by $G$ gauge theory on $M_4\times S^1$ with a twist
around $S^1$ that reduces $G$ to $H$; this further reduces on $M_4$
to $H$ gauge theory with coupling parameter $\tau=iR/S$. The
comparison of these two descriptions is essentially the perspective
offered in \cite{Vafa} on electric-magnetic duality in four
dimensions for non-simply-laced groups.

We can learn more if we do not reduce all the way to four
dimensions. We take $M_6=\R\times S^1\times \TN_k$, where $S^1$ has
circumference $2\pi S$ and the fiber at infinity of
$\vec\mu:\TN_k\to\R^3$ has circumference $2\pi R$.  For small $S$,
we get Description I, which involves $H^\vee$ instantons on $\TN_k$.
For small $R$, we get Description II, which now involves level 1
modules of the affine Kac-Moody algebra of $G$, twisted by
$\lambda$.  This is equivalent to the claim of \cite{BF} (although
the roles of $H$ and $H^\vee$ are reversed in our presentation
here).  To make contact with the formulation given in \cite{BF}, one
must know that the Langlands or GNO dual of the Kac-Moody group of
$H^\vee$ is not the Kac-Moody group of $H$ but the $\lambda$-twisted
Kac-Moody group of $G$.

\remark The construction in the last paragraph does not appear to
have an analog with a twist around the circle at infinity in
$\TN_k$. Precisely because this circle is contractible in the
interior of $\TN_k$, it is not possible to twist by a discrete
symmetry of the $(0,2)$ model in going around this circle.
\bigskip

\subsubsection{Points With Multiplicity}
So far we have considered the points $\vec x_j$ to be distinct, so
that $\TN_k$ is smooth.  It is important, however, to consider the
behavior as some of the $\vec x_j$ become coincident.  In general,
suppose that the $\vec x_j$ become coincident for $j=i_1,\dots,i_r$.
Without any essential change in the following remarks, we could
allow several subsets of $\vec x_j$ to simultaneously become
coincident.

In Description I, when this happens, $\TN_k$ develops a ${\mathrm
A}_{r-1}$ singularity, which is an orbifold singularity, locally
modeled by $\R^4/\Z_r$.  Gauge theory on $\R^4/\Z_r$ is defined as
$\Z_r$-invariant gauge theory on $\R^4$, but the notion of $\Z_r$
invariance depends on the choice of how $\Z_r$ acts on the fiber of
a $G$-bundle at the fixed point (the origin in $\R^4$).  Such a
choice is a homomorphism $\phi:\Z_r\to G$. When a $\Z_r$ orbifold
singularity develops, the space $\V$ of BPS states becomes graded by
the choice of $\phi$.  This is in addition to the grading by the
part of the characteristic class $a$ that can be defined on the
complement of the singularity.

The dual in Description II is that $r$ of the points $\vec x_j$ that
support level 1 holomorphic WZW models become coincident, say at a
point $\vec y\in \R^3$.  Then the submanifold $\R\times S^1\times
\vec y$ of $M_6$ supports a level $r$ holomorphic WZW model of type
$G$. The affine Kac-Moody group of $G$ supports several inequivalent
integrable highest weight modules of level $r$, and when the points
$\vec x_{i_1},\dots, \vec x_{i_r}$ become coincident, the Hilbert
space of the theory decomposes as a direct sum of subspaces
transforming in different such representations. This also gives a
decomposition of the space $\V$ of BPS states.

So duality must establish a correspondence between two types of data
($\phi$ and the relevant part of $a$ on one side; a choice of level
$r$ integrable representation on the other side).
 Such a correspondence is used in \cite{BF} and can be
described as follows in physical terms. For simplicity, we set
$G=E_8$, so that we can dispense with $a$.  Consider
three-dimensional Chern-Simons theory with gauge group $G$ at level
$r$.  A Wilson loop can be considered in any irreducible
representation $R$ that is the highest weight of a level $r$
integrable module of the affine Kac-Moody algebra.  On the other
hand, around the Wilson loop, the gauge field acquires a monodromy
that is an element of $G$ of order $r$.  This gives a correspondence
between level $r$ integrable modules and conjugacy classes of order
$r$.  It should be possible to use string theory arguments to show
that this is the correspondence that enters in comparing
Descriptions I and II, but this will not be attempted here.

The general story is to consider points $\vec y_1,\dots,\vec y_s\in
\R^3$ with multiplicities $r_1,\dots, r_s$.  In our presentation, we
obtain this case starting with $k=\sum_{i=1}^s r_i$ points, all of
multiplicity 1, and letting them coalesce in clumps of the
appropriate sizes.  Then in Description I, we consider a $\TN_k$
space that is constrained to have singularities of type ${\mathrm
A}_{r_i-1}$ for $i=1,\dots,s$.  Gauge theory at the $i^{th}$
singularity is defined by a choice of homomorphisms
$\phi_i:\Z_{r_i}\to G$.  Description II is based on supersymmetric
gauge theory on $\R\times S^1\times \R^3$, with each $\R\times
S^1\times \vec y_i$ supporting a holomorphic level $r_i$ WZW model,
associated with the level $r_i$ integrable representation that
corresponds to $\phi_i$.

This correspondence, moreover, is compatible with further
coalescences of points, in close parallel to the picture of
\cite{BF}.  Any further coalescence of points leads to a further
decomposition of $\V$ on both sides.

\bigskip\noindent{\it Acknowledgments} I thank A. Braverman for explanations of the  ideas of \cite{BF}.
I also thank S. Cherkis for careful comments on the manuscript and
G. Segal for a discussion of the sense in which the $(0,2)$ theory
does not have a Lagrangian.

\bibliographystyle{amsalpha}

\end{document}